\documentclass[aps,prx,onecolumn,citeautoscript,nofootinbib,preprint]{revtex4}
\usepackage{amsmath}
\usepackage{amssymb}
\usepackage{float}
\usepackage[section]{placeins}
\usepackage{bbm}
\usepackage{bm}
\usepackage{comment}
\usepackage{graphicx}
\usepackage{physics}
\usepackage{color}
\usepackage[papersize={8.5in,11in}]{geometry}
\usepackage[colorlinks=true]{hyperref}
\hypersetup{
    bookmarks=true,         
    unicode=false,          
    pdftoolbar=true,        
    pdfmenubar=true,        
    pdffitwindow=false,     
    pdfstartview={FitH},    
    pdftitle={S},    
    pdfauthor={S. Sachdev},     
    pdfsubject={},   
    pdfcreator={},   
    pdfproducer={}, 
    pdfkeywords={} {} {}, 
    pdfnewwindow=true,      
    colorlinks=true,       
    linkcolor=magenta, 
    citecolor=blue,        
    filecolor=magenta,      
    urlcolor=blue           
}

\usepackage{amsfonts}
\usepackage{upgreek}
\usepackage{slashed}
\usepackage{latexsym}
\usepackage[export]{adjustbox}
\usepackage{dsfont}
\usepackage{subcaption}
\begin{document}

\preprint{\href{https://arxiv.org/abs/2205.02285}{arXiv:2205.02285}}
\title{Statistical mechanics of strange metals and black holes}
\author{Subir Sachdev}
\affiliation{Department of Physics, Harvard University, Cambridge MA-02138, USA}
\affiliation{School of Natural Sciences, Institute for Advanced Study, Princeton, NJ-08540, USA}
\affiliation{International Centre for Theoretical Sciences, Tata Institute of Fundamental Research, Bengaluru 560 089, India}
\date{October 7, 2022}
\begin{abstract}
A colloquium style review of the connections between the Sachdev-Ye-Kitaev model and 
strange metals without quasiparticles, and between the SYK model and the quantum properties of black holes.
Along with other insights, this connection has led to an understanding of the universal form of the low energy density of states of charged black holes in asymptotically 3+1 dimensional Minkowski space.
\\~\\
{\sffamily 
\begin{center} Published in {\it \href{https://www.icts.res.in/newsletter}{ICTS News}}, volume 8, issue 1 (2022) \\ Newsletter of the International Centre for Theoretical Sciences, Tata Institute of Fundamental Research\\
This arXiv version (\href{https://arxiv.org/abs/2205.02285}{arXiv:2205.02285}) has been updated with a few additional comments and references.\\~\\
\includegraphics[width=8cm]{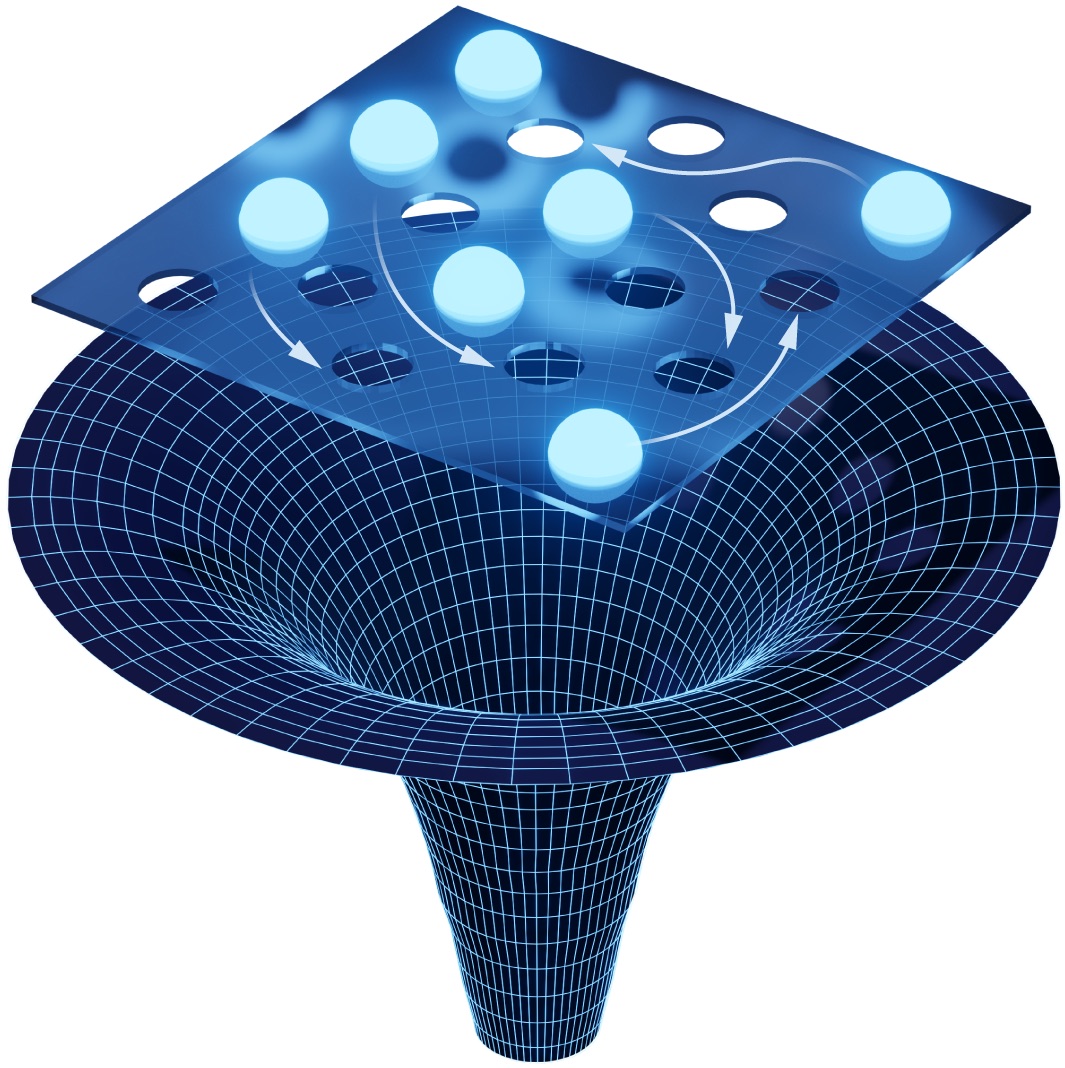}\\
{\tiny Figure by Lucy Reading-Ikkanda, Simons Foundation}
\end{center}
}
\end{abstract}

\pacs{Valid PACS appear here}
\maketitle{}


The meeting of the American Physical Society in New York in 1987 celebrated the discovery of high temperature superconductivity in the cuprate series of compounds. I was fortunate to be present as a young postdoc at this `Woodstock' meeting.
Physicists were confidently applying 
the till-then highly successful methods of condensed matter physics to these materials, expecting a rapid explanation of their remarkable properties. But the ensuing decades have shown that new paradigms of the collective quantum behavior of electrons would be needed, which would take years to develop.
The biggest mystery, as became evident early on, was the unusual metallic state of these materials, above the superconducting critical temperature. This `strange metal' as it has since come to be called, displayed unusual temperature and frequency dependencies in its properties, which indicated that the strange metal was an entangled many-body quantum state without `quasiparticles'. 
Almost all of condensed matter physics is built on the idea of quasiparticles: it allows us to account for the Coulomb interactions between electrons, by assuming their main effect is to renormalize each electron with a cloud of electron-hole pairs, after which we can treat each electron as a nearly independent quasiparticle. This decomposition of the excitations of a many-body system into a composite of simple quasiparticle excitations is an assumption so deeply engrained in the theoretical framework that it is usually left unstated. 

Complex many-particle quantum entanglement is also a central theme in another major puzzle in theoretical physics.
In 1974, Stephen Hawking \cite{Hawking74} combined heretofore distinct pillars of physics: the quantum theory of microscopic particles like the electron, and Einstein's theory of general relativity which applies on astrophysical scales. He argued that the application of quantum theory to the black hole solutions of general relativity led to the remarkable conclusion that each black hole had a non-zero temperature, and an associated entropy (which had been postulated earlier by Bekenstein \cite{Bekenstein73}). Hawking's arguments were based upon semi-classical methods, related to the old quantum theory of Bohr and Sommerfeld. It was not at all clear whether Hawking's results were compatible with the microscopic quantum theory of Schr\"odinger and Heisenberg, which is known today to apply without change to all the microscopic constituents of our universe. Indeed, Hawking famously stated in the early days that perhaps the Schr\"odinger-Heisenberg quantum theory broke down near black holes. Today, there is an emerging consensus that the Schr\"odinger-Heisenberg quantum theory is indeed compatible with general relativity and black holes, and complex and chaotic many-particle quantum entanglement is the key to resolving the difficulties in the semi-classical description.

The last few decades have seen much progress in our understanding of the remarkable physical consequences of many-particle quantum entanglement, coming from a synthesis of ideas from quantum condensed matter theory, quantum information science, quantum field theory, string theory, and also from modern mathematical ideas built on category theory. Here, I will discuss some of the insights that have emerged from a study of the Sachdev-Ye-Kitaev (SYK) model. I proposed a closely related model with the same physical properties in 1992, some of which were described in a paper with my first graduate student, Jinwu Ye \cite{SY}. Alexei Kitaev \cite{kitaev_talk} proposed a modification in 2015 which simplified its solution, and enabled important insights from a more refined analysis \cite{Sachdev15,kitaevsuh,Maldacena_syk,JMDS16b,Fu16,Cotler16,Bagrets17,StanfordWitten,Moitra18,Sachdev19,luca20,matt22}. My motivation in 1992 was to write down the simplest model of a metal without quasiparticles, as a starting point towards addressing the strange metal problem of the cuprates. 
Additional properties of the SYK model were described by Olivier Parcollet and Antoine Georges in Refs.~\cite{Parcollet1,GPS2} in 1999-2001, and in 2010 I pointed out \cite{SS10,Sachdev:2010uj} that the SYK model also provided a remarkable description of the low temperature properties of certain black holes \cite{McGreevy10}. This connection has since undergone rapid development and has been made quite precise. 
The SYK model shows that the quantum entanglement responsible for the absence of quasiparticles in strange metals is closely connected to that needed for a microscopic quantum theory of black holes.

\section{Foundations by Boltzmann}

Let's start by recalling two foundational contributions by Boltzmann to statistical mechanics. 

First, in 1870, Boltzmann gave a precise definition of the  thermodynamic entropy $S$ in statistical terms:
\begin{equation}
S = k_B \ln W\,, \label{b1}
\end{equation}
where $k_B$ is Boltzmann's constant, and $W$ is number of microstates consistent with macroscopically observed properties. The value of $W$ diverges exponentially with the volume of the system, and so $S$ is extensive {\it i.e.\/} proportional to the volume. Boltzmann was thinking in terms of a dilute classical gas of molecules, as found in the atmosphere. But Boltzmann's definition works also for quantum systems, upon replacing $W$ by $D(E)$, the density of the energy eigenstates of the many-body quantum system per unit energy interval; then we have
\begin{equation}
D(E) \sim \exp \left( S(E)/k_B \right)\,, \label{b2}
\end{equation}
where $S(E)$ is the thermodynamic entropy in the microcanonical ensemble with extensive energy $E$.

\begin{figure}
\begin{center}
\includegraphics[scale=0.06]{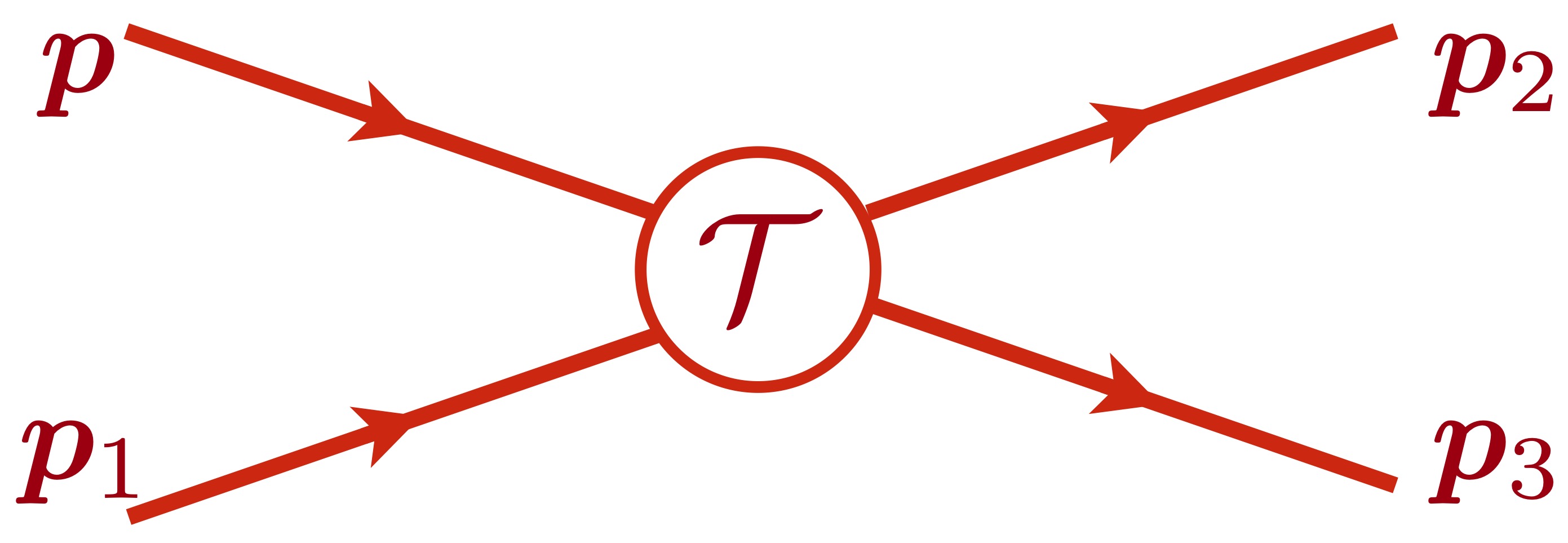}
\end{center}
\caption{Collision between two molecules. The collision term in the Boltzmann equation is proportional to the absolute square of the ${T}$-matrix.}
\label{fig1}
\end{figure}

Second, in 1872, Boltzmann's equation gave a correct description of the time evolution of the observable properties of a dilute gas in response to external forces. He applied Newton's laws of motion to individual molecules, and obtained an equation for $f_{\bm p}$, the density of particles with momentum ${\bm p}$. In a spatially uniform situation, Boltzmann's equation takes the form
\begin{equation}
\frac{\partial f_{\bm p}}{\partial t} + {\bm F} \cdot \nabla_{\bm p} f_{\bm p} = \mathcal{C}[f]\,, \label{be}
\end{equation}
where $t$ is time, and ${\bm F}$ is the external force. The left-hand-side of (\ref{be}) is just a restatement of Newton's laws for individual molecules. Boltzmann's innovation was the right-hand-side, which describes collisions between the molecules. Boltzmann introduced the concept of `molecular chaos', which asserted that in a sufficiently dilute gas successive collisions were statistically independent. With this assumption, Boltzmann showed that
\begin{equation}
\mathcal{C}[f] \propto \int_{{\bm p}_{1,2,3}} \cdots \left [f_{\bm p} f_{{\bm p}_1} - f_{{\bm p}_2} f_{{\bm p}_3} \right] \label{coll1}
\end{equation}
for a collision between molecules as shown in Fig.~\ref{fig1}. The statistical independence of collisions is reflected in the products of the densities in (\ref{coll1}), and the second term represents the time-reversed collision.

\section{Ordinary and strange metals}

The remarkable fact is that Boltzmann's equation also applies, with relatively minor modifications, in situations very different from the dilute classical gas: it also applies to the dense quantum gas of electrons found in ordinary metals. Now collisions become rare because of Pauli's exclusion principle, and the statistical independence of collisions is assumed to continue to apply. The main modification is that the collision term in (\ref{coll1}) is replaced by
\begin{equation}
\mathcal{C}[f] \propto \int_{{\bm p}_{1,2,3}} \cdots \left [f_{\bm p} f_{{\bm p}_1}(1-f_{{\bm p}_2})(1-f_{{\bm p}_3}) - f_{{\bm p}_2} f_{{\bm p}_3} (1-f_{{\bm p}})(1-f_{{\bm p}_1})\right] \,,\label{coll2}
\end{equation}
where the additional $(1-f)$ factors ensure that the final states of collisions are not occupied. Now the $f_{\bm p}$ measure the distribution of electronic quasiparticles, and the cloud of particle-hole pairs around each electron only renormalizes the microscopic scattering cross-section.
Such a quantum Boltzmann equation is the foundation of the quasiparticle theory of the electron gas in metals, superconductors, semiconductors, and insulators, and indeed almost all of condensed matter physics before the 1980's. One of its important predictions is that as temperature $T \rightarrow 0$, the typical time between collisions, $t_c$, diverges as $t_c \sim 1/T^2$.

One can now ask, how short can we make $t_c$ before we cannot ignore the quantum interference between successive collisions, and the concept of quasiparticles does not make sense? An energy-time uncertainty-principle argument indicates that {\it any\/} many-body quantum system should have a relaxation time \cite{ssbook}
\begin{equation}
t_r \geq \alpha \frac{\hbar}{k_B T} \quad, \quad T \rightarrow 0\,, \label{taueq}
\end{equation}
where $\alpha$ is a dimensionless, $T$-independent constant. For systems with quasiparticles, we expect $t_c \sim t_r$, and we have introduced a general relaxation time $t_r$ to allow a more general discussion in systems without quasiparticles. From studies of various quantum critical systems, it was argued \cite{ssbook} that the inequality in (\ref{taueq}) becomes an equality when quasiparticles are absent, as in strange metals. Recent experiments \cite{admr20} on the strange metal in cuprate superconductors have measured a particular relaxation time by connecting it to the angle dependence of the resistivity in a magnetic field, and indeed found it obeys (\ref{taueq}) as an equality, with $\alpha \approx 1.2$. This is often stated as the strange metal exhibiting `Planckian time' dynamics \cite{Hartnoll21}.

\section{Quantum black holes and holography}

\begin{figure}
\begin{center}
\includegraphics[scale=0.07]{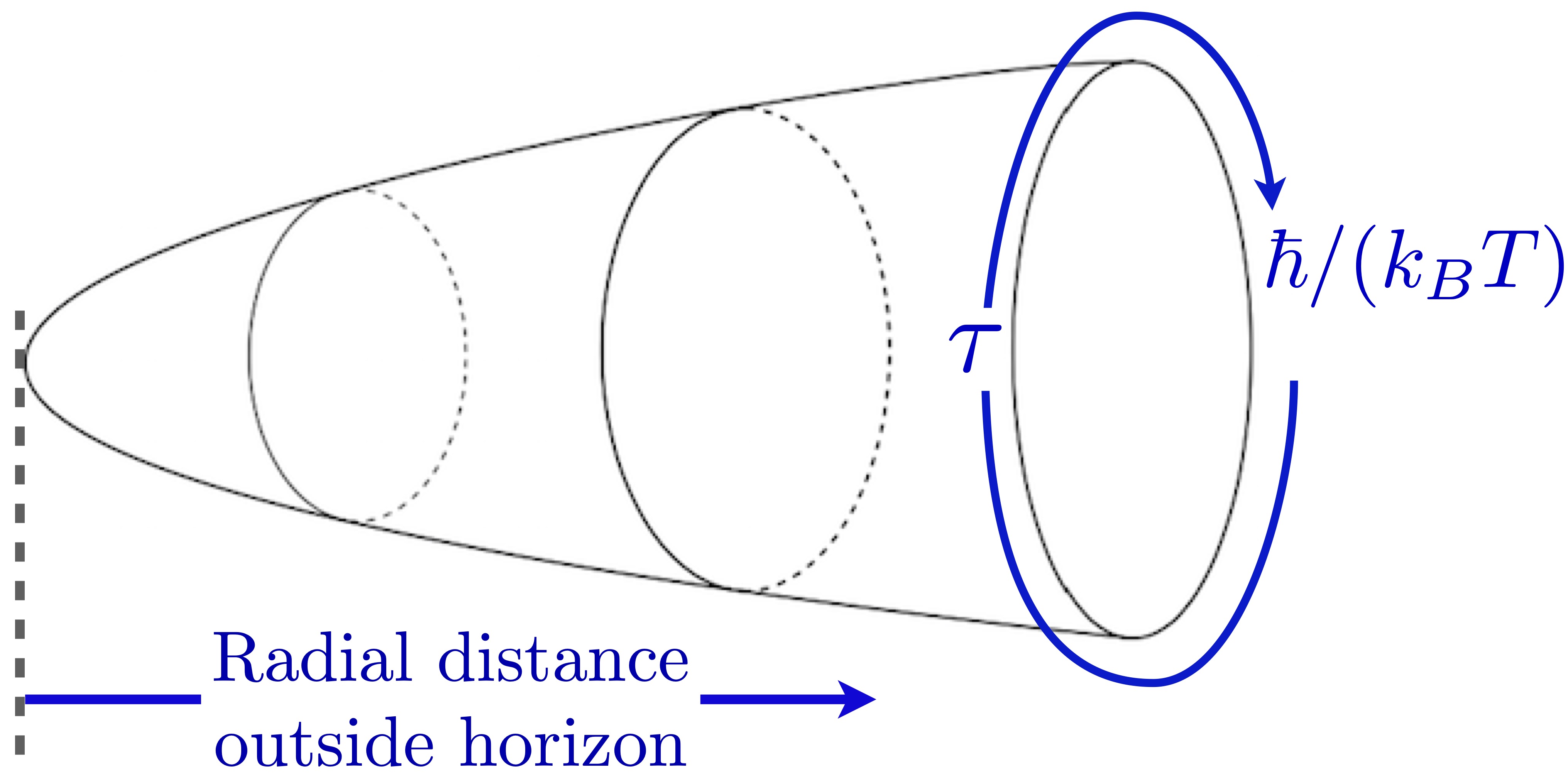}
\end{center}
\caption{Spacetime geometry outside a black hole. Only the radial direction and the imaginary time direction $\tau$ are shown, and the two angular directions are not shown.}
\label{fig2}
\end{figure}

We can write the quantum theory of black holes as a Feynman path integral over the spacetime metric $g_{\mu\nu}$, and the electromagnetic gauge field $a_\mu$: this involves computing the partition function
\begin{equation}
\mathcal{Z} = \int \mathcal{D} g_{\mu \nu} \mathcal{D} a_\mu \exp \left( - \frac{1}{\hbar} \int d^d x \int_0^{\hbar/(k_B T)} \!\!\!\!\!\! d \tau 
\sqrt{g} \, \mathcal{L}_d [ g_{\mu\nu}, a_\mu ] \right) \label{feyn}
\end{equation}
over fields in $(d+1)$-dimensional spacetime, with $\mathcal{L}_d$ the Lagrangian of classical Einstein-Maxwell theory in $d+1$ spacetime dimensions, and $g$ the determinant of the metric. Here $\tau$ is time analytically continued to the imaginary axis, which is taken to lie on a circle of circumference $\hbar/(k_B T)$, where $\hbar$ is Planck's constant. 
This constraint on imaginary time follows from the correspondence between the evolution operator ${U}(t)$ for real time $t$ in quantum mechanics, and the Boltzmann-Gibbs partition function $\mathcal{Z}$ for a quantum system with Hamiltonian $\mathcal{H}$:
\begin{equation}
{U} (t)  = \exp \left( - i \mathcal{H} t/ \hbar \right) \quad \Leftrightarrow \quad \mathcal{Z} = \mbox{Tr} \exp \left( - \mathcal{H}/(k_B T) \right)\,.
\end{equation}
In imaginary time, the spacetime geometry outside a black hole is that of a `cigar' as shown in Fig.~\ref{fig2}.
Note that all dependence of (\ref{feyn}) on $\hbar$ and $T$ is explicit, and there is no $\hbar$ or $T$ in the Lagrangian $\mathcal{L}_d$.

Formally, the path integral in (\ref{feyn}) is pathological because it includes infinities that cannot be controlled by the usual renormalization tricks of quantum field theory. Nevertheless, Gibbons and Hawking \cite{Gibbons_Hawking} boldly decided to evaluate it in the semi-classical limit, where one only includes the contribution of the cigar saddle point in Fig.~\ref{fig2}. For a black hole, they also imposed the requirement that spacetime was smooth at the horizon in imaginary time. From this relatively simple computation, they were able to obtain the thermodynamic properties of a black hole, including its temperature and entropy. For a neutral black hole of mass $M$ in $d=3$ they found
\begin{equation}
\frac{S}{k_B} =  \frac{c^3 A}{4 \hbar G} \quad, \quad \frac{k_B T}{\hbar} = \frac{c^3}{8 \pi G M} \label{ST}
\end{equation}
where $c$ is the velocity of light, $G$ is Newton's gravitational constant, and $A = 4 \pi R^2$ is the {\it area} of the black hole horizon with $R = 2 GM/c^2$ the horizon radius.

The revolutionary results in (\ref{ST}) raised many more questions than they answered. Is this semi-classical computation of thermodynamics compatible with Boltzmann's fundamental statistical interpretation of entropy in (\ref{b2})? How does a computation in imaginary time outside a black hole know about the entropy of quantum degrees of freedom inside a black hole? Can one compute the energy eigenvalues of a quantum Hamiltonian describing the inside of a black hole whose density of states $D(E)$ yields a $S(E)$ that is consistent with (\ref{ST}), and the partition function $\mathcal{Z}$ in (\ref{feyn})? With the energy $E$ shifted so that $E=0$ for the ground state, $\mathcal{Z}$ is related to $D(E)$ by
\begin{equation}
\mathcal{Z} = \int_{0^-}^\infty dE D(E) \exp\left( - \frac{E}{k_B T} \right) \, . \label{ZD}
\end{equation}
Many other questions are raised when one considers the fate of the black hole as it evaporates while emitting blackbody radiation at the temperature in (\ref{ST}), and computes the entanglement entropy of the Hawking radiation.

\begin{figure}
\begin{center}
\includegraphics[scale=0.1]{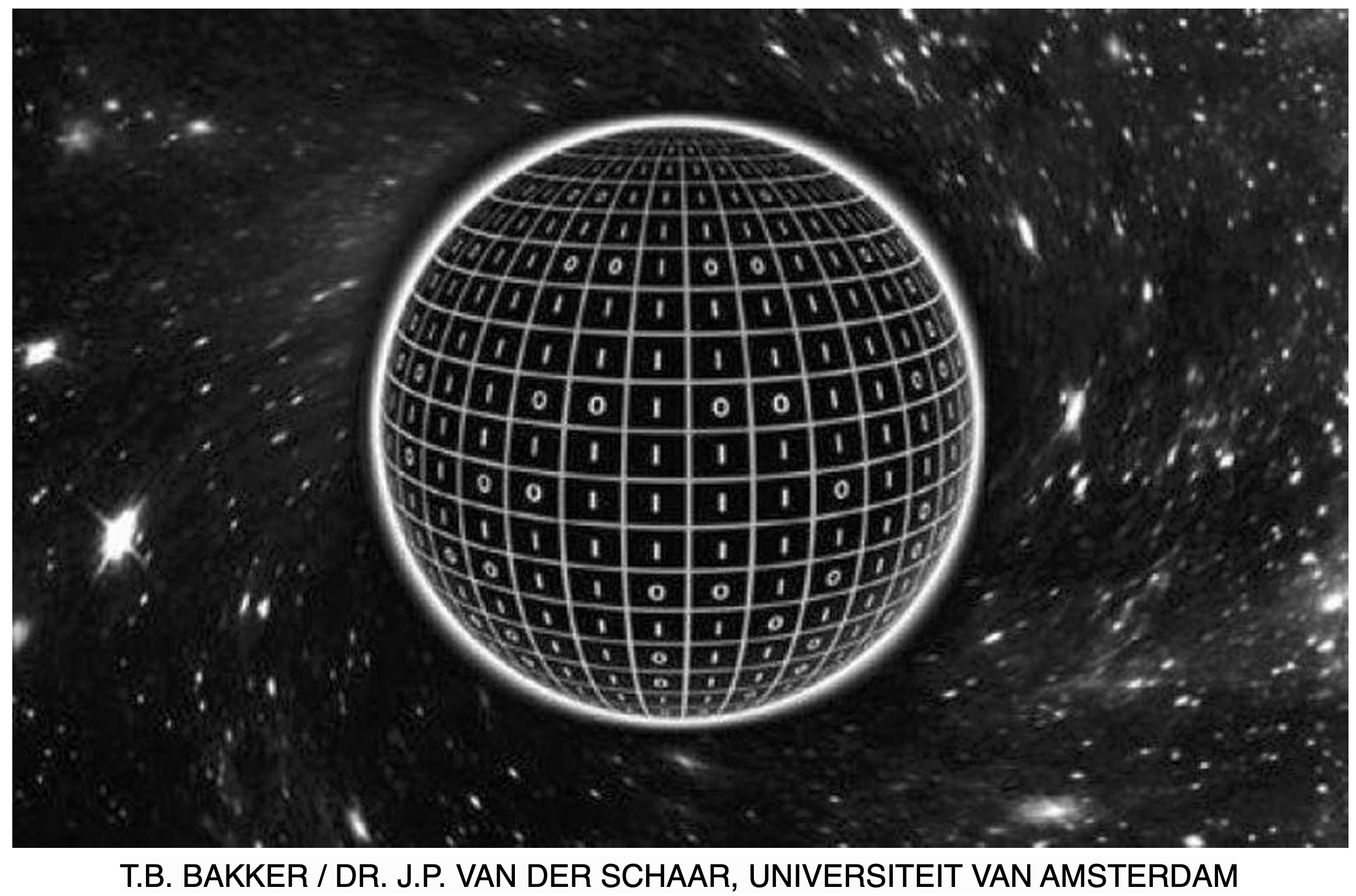}
\end{center}
\caption{Holography: the number of qubits required for the quantum simulation of a black hole is proportional to its surface area.}
\label{fig3}
\end{figure}

A remarkable feature of the entropy in (\ref{ST}) is that it is proportional to the surface {\it area\/} of the black hole. This contrasts with extensive volume proportionality of the entropy, mentioned below (\ref{b1}), obeyed by all other quantum systems. Attempts to understand this feature led to the idea of holography \cite{tHooft,Susskind,Maldacena} illustrated in Fig.~\ref{fig3}. 
Let us try to build a quantum simulation of a black hole by simple two-level systems {\it i.e.\/} qubits. How many qubits will we need? As $N$ qubits can describe $2^N$ linearly-independent quantum states, 
the formula in (\ref{b2}) tells us that the number of qubits is proportional to the entropy, and hence the area of the black hole. So the qubits realize a many-body quantum system which we can imagine residing on its surface {\it i.e.\/} the qubits are a faithful $(d-1)$-dimensional hologram of the complete quantum gravitational theory of the black hole in $d$ spatial dimensions.  

A specific realization of a quantum simulation was found in string theory for `extremal' black holes (defined below) with low energy supersymmetry \cite{David02}. This realization has a ground state with an exponentially large degeneracy, and special features of the supersymmetry were employed to compute this degeneracy, yielding
\begin{equation}
D(E) = \left[\exp (S(E)/k_B)\right] \delta(E) + \ldots\,. \label{Ddelta}
\end{equation}
where $\ldots$ refers to a continuum above an energy gap.
The value of $S(0)$ was found to be precisely that in the Hawking formula in (\ref{ST}) \cite{Strominger96}. However, the zero energy delta-function in (\ref{Ddelta}) is known \cite{matt22,luca22b} to be a special feature of theories with low energy supersymmetry, and is not a property of the generic semi-classical path integral over Einstein gravity in (\ref{feyn}), as we will discuss below (see Fig.~\ref{fig5}).

To move beyond supersymmetric string theory, we ask if there are any general constraints that must be obeyed by the many-body system realized by the interactions between the qubits. An important constraint comes from an earlier result by Vishveshwara \cite{cvc}. He computed the relaxation time, $t_r$ of quasi-normal modes of black holes in Einstein's classical theory; this is the time in which a black hole relaxes exponentially back to a spherical shape after it has been perturbed by another body:  
\begin{equation}
t_r = \alpha' \frac{8 \pi G M }{c^3} \,,\label{tr}
\end{equation}
where $\alpha'$ is a numerical constant of order unity dependent upon the precise quasi-normal mode.
Comparing Vishveshwara's result in (\ref{tr}) with Hawking's result in  (\ref{ST}), we can write
\begin{equation}
t_r = \alpha' \frac{\hbar}{k_B T} \label{tra}
\end{equation}
which is exactly of the form in (\ref{taueq}) for many body quantum systems without quasiparticles! This is a key hint that the holographic quantum simulation of a black hole must involve a quantum system without quasiparticle excitations, if it is reproduce basic known features of black hole dynamics. At this point, it is interesting to note that measurements of $t_r$ in binary black hole mergers by LIGO-Virgo \cite{BHbound} do indeed fall around the value of $\hbar/(k_B T)$.

\section{The SYK model}

\begin{figure}
\begin{center}
\includegraphics[scale=0.055]{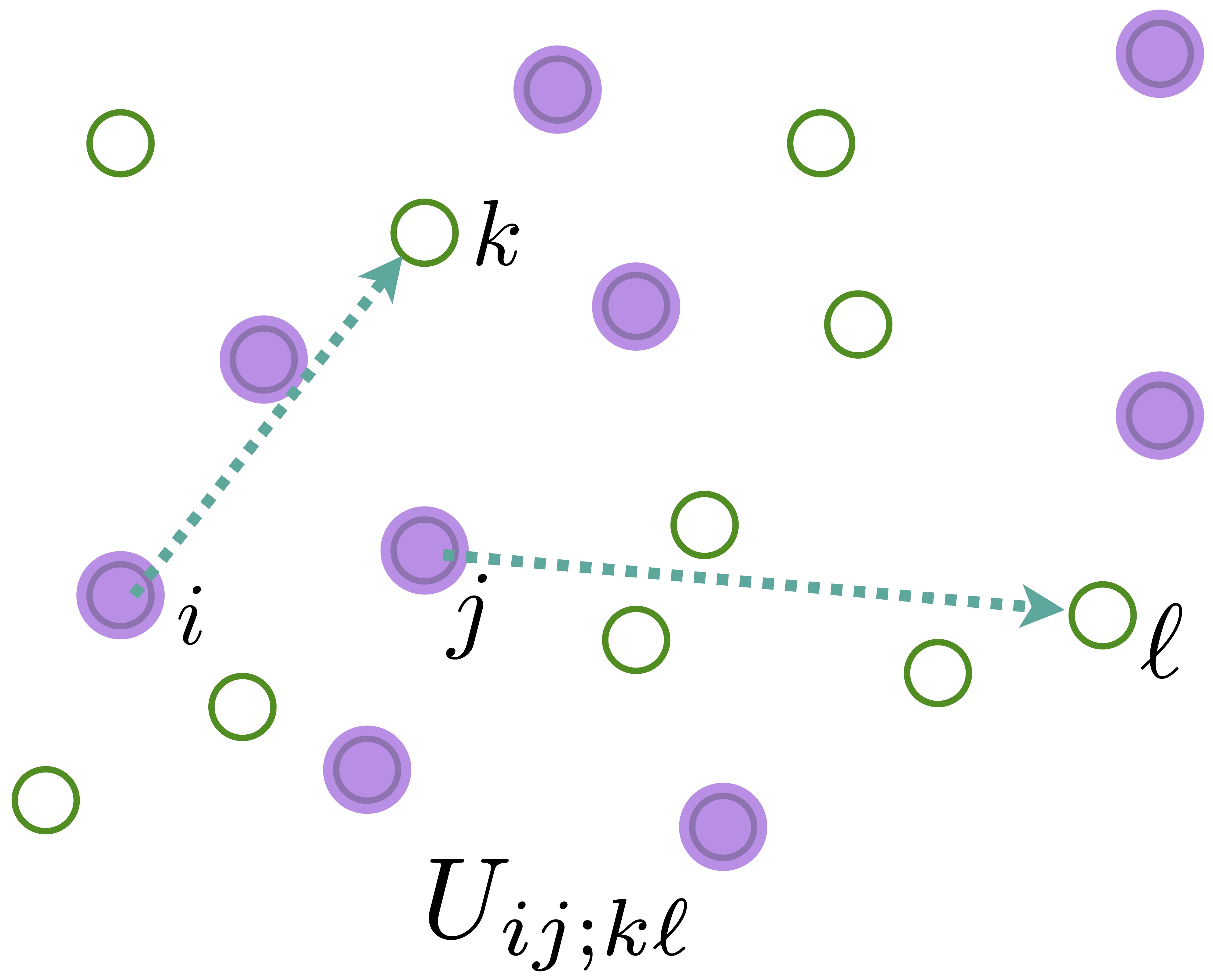}
\end{center}
\caption{The SYK model: fermions undergo the transition (`collision') shown with quantum amplitude $U_{ij;k\ell}$.}
\label{fig4}
\end{figure}

The Hamiltonian of a version of a SYK model is illustrated in Fig.~\ref{fig4}. We take a system with fermions $\psi_i$, $i=1\ldots N$ states. Depending upon physical realizations, the label $i$ could be position or an orbital, and it is best to just think of it as an abstract label of a fermionic qubit with the two states $\left|0 \right\rangle$ and $\psi_i^\dagger \left|0 \right\rangle$. We now place ${Q} N$ fermions in these states, so that a density ${Q} \approx 1/2$ is occupied, as shown in Fig.~\ref{fig4}. 
The quantum dynamics is restricted to {\it only\/} have a `collision' term between the fermions, analogous to the right-hand-side of the Boltzmann equation. However, in stark contrast to the Boltzmann equation, we will not make the assumption of statistically independent collisions, and will account for quantum interference between successive collisions: this is the key to building up a many-body state with non-trivial entanglement. So a collision in which fermions move from sites $i$ and $j$ to sites $k$ and $\ell$ is characterized not by a probability, but by a quantum amplitude $U_{ij;k\ell}$, which is a complex number.

The model so defined has a Hilbert space of order $2^N$ states, and a Hamiltonian determined by order $N^4$ numbers $U_{ij;k\ell}$. Determining the spectrum or dynamics of such a Hamiltonian for large $N$ seems like an impossibly formidable task. But if we now make the assumption that the $U_{ij;k\ell}$ are statistically independent random numbers, remarkable progress is possible. Note that we are not considering an ensemble of SYK models with different $U_{ij;k\ell}$, but a single fixed set of $U_{ij;k\ell}$. Most physical properties of this model are self-averaging at large $N$, and so as a technical tool, we can rapidly obtain them by computations on an ensemble of random $U_{ij;k\ell}$. In any case, the analytic results we now describe have been checked by numerical computations on a computer for a fixed set of $U_{ij;k\ell}$.
We recall that even for the Boltzmann equation, there was an ensemble average over the initial positions and momenta of the molecules that was implicitly performed.

Using these methods, key properties of the SYK model have been established (for complete references to the literature, please see the review in Ref.~\cite{SYKRMP}):
\begin{itemize} 
\item There are no quasiparticle excitations, and it exhibits quantum dynamics with a Planckian relaxation time obeying (\ref{tra}) at $T \ll U$, where $U/N^{3/2}$ is the root-mean-square value of the $U_{ijk\ell}$. In particular, the relaxation time is {\it independent\/} of $U$, a feature not present in any ordinary metal with quasiparticles. 
\item At large $N$, the many-body density of states at fixed $Q$ is (see Fig.~\ref{fig5}a)
\begin{equation}
D(E) \sim \frac{1}{N} \exp (N s_0) \sinh \left( \sqrt{2 N \gamma E} \right)\,. \label{de}
\end{equation}
Here $s_0$ is a universal number dependent only on ${Q}$ ($s_0 = 0.4648476991708051 \ldots$ for ${Q}=1/2$), $\gamma \sim 1/U$ is the only parameter dependent upon the strength of the interactions, and the $N$ dependence of the pre-factor is discussed in Ref.~\cite{GKST}.
\begin{figure}
\begin{center}
\includegraphics[width=6in]{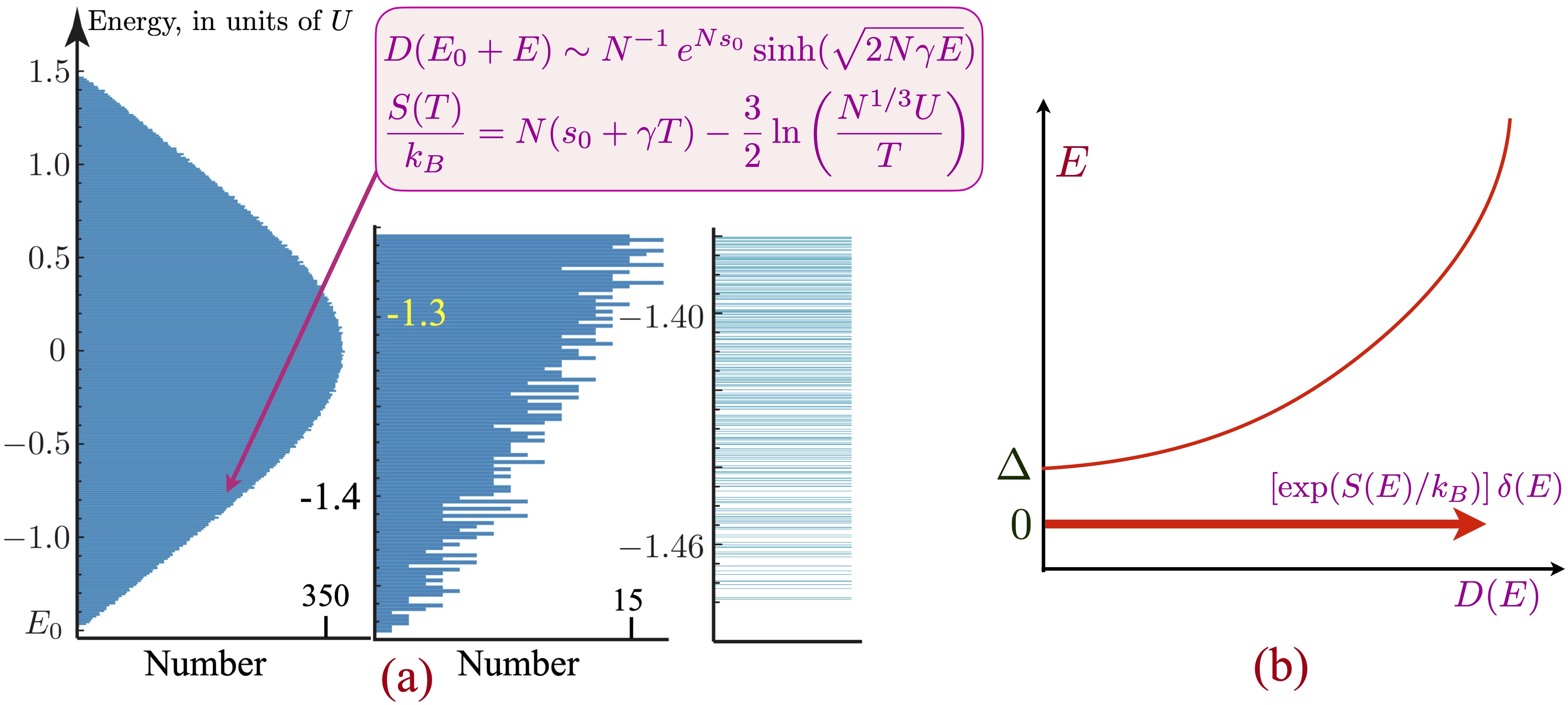}
\end{center}
\caption{(a) Plot of the 65536 many-body eigenvalues of a $N = 32$ Majorana SYK Hamiltonian; however, the analytical results quoted here are for the SYK model with complex fermions which has a similar spectrum. The coarse-grained low energy and 
low temperature behavior is described by (\ref{de}) and (\ref{SSYK}). The lower energy part of this density of states is argued to 
apply to the semi-classical path integral in (\ref{feyn}) for charged black holes at low $T$ ({\it i.e.\/} extremal black holes).
(b) Schematic of the lower energy density of states of the supersymmetric string theory description of extremal black holes 
\cite{matt22} in (\ref{Ddelta}). 
The energy gap $\Delta$ is proportional to the inverse of $S(E=0)$.}
\label{fig5}
\end{figure}
Given $D(E)$, we can compute the partition function from (\ref{ZD}) at a temperature $T$, and hence the low $T$ dependence of the entropy at fixed $Q$
\begin{equation}
\frac{S(T)}{k_B} = N(s_0 + \gamma \, k_B T) - \frac{3}{2}\ln \left(\frac{U}{k_B T} \right) - \frac{\ln N}{2} + \ldots \,. \label{SSYK}
\end{equation}
The limit $\lim_{T \rightarrow 0} \lim_{N \rightarrow \infty} S(T)/(k_B N) = s_0$ is non-zero, implying an energy level spacing exponentially small in $N$ near the ground state. This is very different from systems with quasiparticle excitations, whose energy level spacing vanishes with a positive power of $1/N$ near the ground state. However, there is no exponentially large degeneracy of the ground state itself in the SYK model, unlike the ground states of the string theory solutions leading to (\ref{Ddelta}), and the ground states in Pauling's model of ice \cite{Pauling}. Obtaining the ground state degeneracy requires the opposite order of limits between $T$ and $N$, and numerical studies show that the entropy density does vanish in such a limit for the SYK model. The density of states (\ref{de}) implies that any small energy interval near the ground state contains an exponentially large number of energy eigenstates with an exponentially small spacing in energy (see Fig.~\ref{fig5}a). The wavefunctions of these eigenstates in Fock space change chaotically from one state to the next, providing a realization of maximal many-body quantum chaos \cite{Maldacena16} in a precise sense.
This structure of eigenstates is very different from systems with quasiparticles, for which the lowest energy eigenstates differ only by adding and removing a few quasiparticles. 
\item
The $E$ dependence of the density of states in (\ref{de}) is associated with a time reparameterization mode, and (\ref{de}) shows that its effects are important when $E \sim 1/N$. We can express the low energy quantum fluctuations in terms of a path integral which reparameterizes imaginary time $\tau \rightarrow f(\tau)$, in a manner analogous to the quantum theory of gravity being expressed in terms of the fluctuations of the spacetime metric. There are also quantum fluctuations of a phase mode $\phi (\tau)$, whose time derivative is the charge density, and so we have the partition function
\begin{equation}
\mathcal{Z}_{SYK} = e^{N s_0} \int \mathcal{D} f \mathcal{D} \phi \exp \left( - \frac{1}{\hbar} \int_0^{\hbar/(k_B T)}\!\!\!\!\!\!  d \tau \, \mathcal{L}_{SYK} [ f,\phi] \right) \label{feynsyk}
\end{equation}
The Lagrangian $\mathcal{L}_{SYK}$ is known, and involves a Schwarzian of $f(\tau)$. Remarkably, despite its non-quadratic Lagrangian, the path integral in (\ref{feynsyk}) can be performed exactly \cite{StanfordWitten}, and leads to (\ref{de}).
\end{itemize}

\section{From the SYK model to black holes}

Can we use insights from the path integral over time reparameterizations of the SYK model in (\ref{feynsyk}) to evaluate the path integral over spacetime metrics of black holes in (\ref{feyn})? Remarkably, for a black hole with a non-zero fixed total charge ${Q}$, the answer is yes (for complete references to the literature in the discussion below,
please see the review in Ref.~\cite{SYKRMP}). 

\begin{figure}
\begin{center}
\includegraphics[scale=0.055]{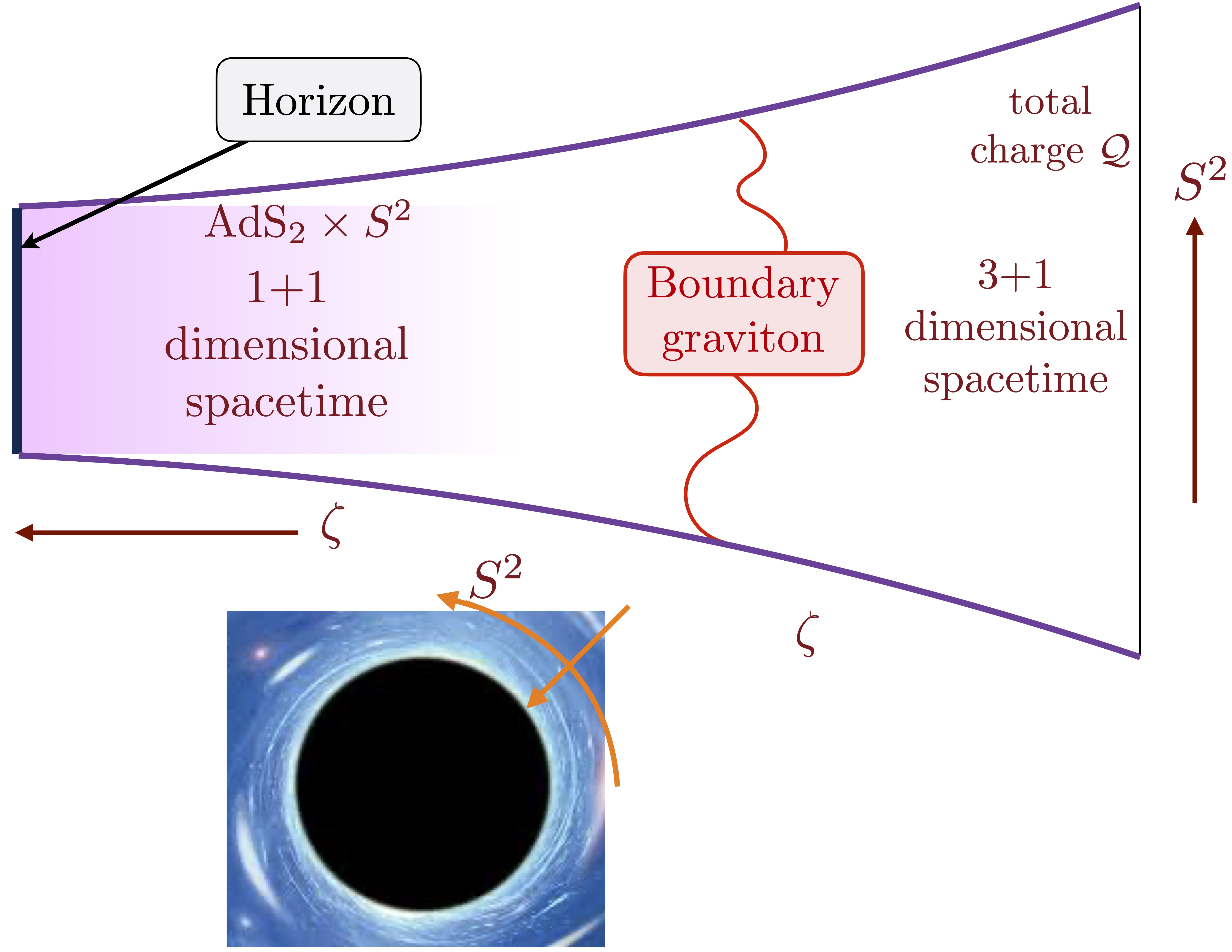}
\end{center}
\caption{Two views of the spacetime outside a charged black hole. $S^2$ is the sphere, and AdS$_2$ is anti-de Sitter space with metric $ds^2 = (d\zeta^2 + d\tau^2)/\zeta^2$.}
\label{fig6}
\end{figure}

The saddle-point solution of the Einstein-Maxwell action for a charged black hole has the form shown in Fig.~\ref{fig6}: while the spacetime is 3+1 dimensional flat Minkowski far from the black hole, it factorizes into a 1+1 dimensional spacetime involving the radial direction $\zeta$, and a 2-dimensional space of non-zero angular momentum modes around the spherical black hole. 
As the black hole temperature $T \rightarrow 0$ (also known as the extremal limit), the non-zero angular momentum modes become unimportant, and we can write the partition function of the charged black hole purely as a theory of quantum gravity in 1+1 spacetime dimensions, which is an extension of a theory known as Jackiw-Teitelboim (JT) gravity; then (\ref{feyn}) reduces to
\begin{equation}
\mathcal{Z}_{JT} = e^{A_0 c^3/(4 \hbar G)} \int \mathcal{D} g_{\mu \nu} \mathcal{D} a_\mu \exp \left( - \frac{1}{\hbar} \int d \zeta \int_0^{\hbar/(k_B T)} \!\!\!\!\!\! d \tau 
\sqrt{g} \, \mathcal{L}_1 [ g_{\mu\nu}, a_\mu ] \right)\,, \label{feynQ}
\end{equation}
where $A_0 = 2 G Q^2/c^4$ is the area of the black hole horizon at $T=0$. The $(1+1)$-dimensional spacetime saddle point of $\mathcal{Z}_{JT}$ has a uniform negative curvature: it is the anti-de Sitter space AdS$_2$, noted in Fig.~\ref{fig6}. Quantum gravity in 1+1 dimensions is especially simple because there is  no graviton, and it is possible to make an explicit holographic mapping to a quantum system in 0+1 dimensions. It turns out that the holographic quantum realization of the 1+1 dimensional theory $\mathcal{Z}_{JT}$ in (\ref{feynQ}) is exactly the $0+1$ dimensional SYK model partition function in $\mathcal{Z}_{SYK}$ in (\ref{feynsyk}). The fluctuations of the metric in the boundary region between the 1+1 dimensional and 3+1 dimensional spacetimes (denoted `boundary graviton' in Fig.~\ref{fig6}) 
are described by the time reparameterization $f(\tau)$, and the boundary value of $a_\mu$ becomes the phase field $\phi (\tau)$. This powerful connection enables us to proceed beyond the semi-classical results of Hawking in (\ref{ST}) for black holes with non-zero charge ${Q}$. Applying this mapping from $\mathcal{Z}_{JT}$ to $\mathcal{Z}_{SYK}$, we obtain a density of states $D(E)$ with precisely the $E$ dependence in (\ref{de}), which also corresponds to that shown for the SYK model in Fig.~\ref{fig5}a.
This should be contrasted with the supersymmetric result in (\ref{Ddelta}) and Fig.~\ref{fig5}b \cite{luca20,matt22}. The parameters in the black hole $D(E)$ can be deduced by comparing the SYK entropy in (\ref{SSYK}) with the low $T$ limit of 
the entropy of a charged black hole in asymptotically 3+1 dimensional Minkowski spacetime, which is
\begin{equation}
\frac{S(T)}{k_B} = \frac{c^3}{4 \hbar G}\left(A_0 + 2\sqrt{\pi} A_0^{3/2} \frac{ k_B T}{\hbar c} \right) - \frac{3}{2}  \ln \left( \frac{(\hbar c^5/G)^{1/2}}{k_B T} \right) + \ldots\,. \label{SG}
\end{equation}
The first two terms in (\ref{SG}) correspond to Hawking's area law in (\ref{ST}), with an horizon area $A$ which increases linearly with $T$ from its $T=0$ value $A_0$. The linear $T$ dependence can also be obtained from the saddle-point action of $\mathcal{Z}_{JT}$ in (\ref{feynQ}). These first two terms in (\ref{SG}) are  in one-to-one correspondence with the first two terms in (\ref{SSYK}) for the SYK model. Indeed the correspondence extends also to the third terms in 
(\ref{SG}) and (\ref{SSYK}), both of which are obtained by performing the equivalent path integrals in $\mathcal{Z}_{JT}$ and $\mathcal{Z}_{SYK}$.
These terms are obtained in an expansion in small $\hbar G$ for the black hole, in place of large $N$ for the SYK model. 
There is also a $T$-independent term, $(-559/180) \ln (A_0 c^3/(\hbar G))$, arising from other massless modes of the $3+1$ dimensional black hole, not included in (\ref{SG}) \cite{Sen12,luca20,luca22a}; this analogous to the $\ln N$ term in (\ref{SSYK}), but the co-efficients of the logarithms differ because the matter content of the two theories differ.
In this manner, we obtain the expression for the universal form of the low energy density of states of charged black holes in asymptotically 3+1 dimensional Minkowski spacetime, which is the analog of (\ref{de}) for the SYK model
\begin{equation}
D(E) \sim \left( \frac{A_0 c^3}{\hbar G} \right)^{-347/90} \exp\left( \frac{A_0 c^3}{4 \hbar G} \right) \sinh \left( \left[\sqrt{\pi} A_0^{3/2} \frac{c^3}{\hbar G} \frac{E}{\hbar c} \right]^{1/2} \right) \,. \label{DEF}
\end{equation}
This is very different from the result in (\ref{Ddelta}) obtained from string theory. Recall that $A_0$ is the area of the black hole horizon at $T=0$; all other parameters in (\ref{DEF}) are fundamental constants of nature. The result in (\ref{DEF}) is a rare formula which combines Planck's constant $\hbar$ with Newton's gravitational constant $G$: the exponential prefactor was obtained by Hawking, and the sinh follows from developments ensuing from the solution of the SYK model. In particular, the time reparameterization mode is important when the sinh in (\ref{DEF}) becomes of order unity, and such corrections address \cite{luca20,matt22} issues raised in Ref.~\cite{Preskill91}.

We have now answered one of the questions raised by Hawking's semi-classical computation of black hole entropy, at least for the case of a charged black hole at low $T$. There is a perfectly well-defined quantum system, with precise and discrete energy levels, whose many body density of states $D(E)$ is described at low energies by a gravitational theory expressed in terms of the semi-classical fluctuations of a spacetime metric: this is the SYK model. This connection has enabled computation of the logarithmic correction to black hole entropy in (\ref{SG}) and the density of states in (\ref{DEF}), and has led to some understanding of the key role played by many-body quantum chaos in the black hole microstates \cite{Shenker13}. The random couplings in the SYK model mainly play the role of a powerful computational tool for accessing the physics of chaotic behavior, as was also the case in single-particle problems such as `quantum billiards' \cite{Bohigas}. The quantities being studied here self-average in a single realization of the random couplings, and are not sensitive to the particular member chosen from the random ensemble. It is also worth noting here that the Boltzmann equation also involves an implicit average over an ensemble of initial conditions for the quasiparticles. 

The connection between the SYK model and black holes does {\it not\/} imply that the ultimate high energy and short distance physics is described by the SYK model. That likely requires string theory, in which supersymmetry is restored at high energies. Nevertheless, the SYK model provides a much simpler quantum simulation of the low energy physics in certain cases, including the complex quantum entanglement and the maximal many-body quantum chaos. The semi-classical path integral over Einstein gravity can capture certain coarse-grained properties of the underlying Schr\"odinger-Heisenberg quantum theory, and is not sensitive to all the microscopic details at the smallest length scales. Recent work \cite{Almheiri20,Bousso} has shown that the path integral over Einstein gravity can also consistently describe the time evolution of the entanglement entropy of an evaporating black hole, provided we also include the contributions of `wormhole' solutions of Einstein gravity. Such wormholes can also be realized by the quantum states of the SYK model.
The idea that Einstein quantum gravity is a coarse-grained description of the microscopics has also led to many theoretical advances by employing averages over ensembles of consistent, microscopic, quantum theories \cite{Chandra}.

\section{From the SYK model to strange metals}

To conclude, we return to the original motivation for studying the SYK model, to the strange metal phase of the cuprate superconductors and related compounds. Realistic models of the lattice scale physics of these materials involve mobile electrons with strong interactions. These models can be solved by methods closely related to those of the SYK model, after they are extended to large spatial dimension $d$, and upon including random spin exchange interactions. These large $d$ models display strange metal phases with many similarities to observations \cite{SYKRMP}. In particular, such strange metals do have a time reparameterization low energy mode, and this mode leads to a linear-in-temperature resistivity at the smallest $T$ \cite{guo2020}, as is observed in the strange metals of the cuprates. 

Other works have focused on the strange metal directly in $d=2$ dimensions, which is the physically relevant dimension for the cuprates. 
Rather than departing from the SYK model, physically realistic models of such strange metals can be built from `Yukawa-SYK' models of fermions and bosons with Yukawa couplings \cite{SYKRMP}.
An important consideration is the role of the Fermi surface: for free electrons, this is the surface which divides occupied and empty electron states in momentum space. A suitably defined Fermi surface is also present when there are interactions between the electrons in an ordinary metal, and also when quasiparticles are no longer present in a strange metal.  An interesting and much studied strange metal is obtained when the interactions between the Fermi surface excitations are mediated by the exchange of a boson with a gapless energy spectrum. Such a boson may be the order parameter of a symmetry-breaking quantum phase transition, or an emergent gauge field in a correlated metal. However, this strange metal has a zero resistance because the low energy theory has a continuous translation symmetry. But recent studies \cite{Altman1,Patel1,Guo2022,Patel2} have shown that adding a SYK-like randomness to the boson-electron Yukawa coupling does lead to a linear-in-temperature resistivity as $T \rightarrow 0$. Such randomness can arise from impurities in the strange metal, and quantifying the role of randomness is an important direction for the further study of strange metals.

\subsection*{Acknowledgements} I thank Matthew Heydeman, G. Joaquin Turiaci and Spenta Wadia for important comments on the manuscript. This research has been supported by the U.S. National Science Foundation, most recently under grant No. DMR-2002850.

\bibliography{refs}

\begin{thebibliography}{50}%
\makeatletter
\providecommand \@ifxundefined [1]{%
 \@ifx{#1\undefined}
}%
\providecommand \@ifnum [1]{%
 \ifnum #1\expandafter \@firstoftwo
 \else \expandafter \@secondoftwo
 \fi
}%
\providecommand \@ifx [1]{%
 \ifx #1\expandafter \@firstoftwo
 \else \expandafter \@secondoftwo
 \fi
}%
\providecommand \natexlab [1]{#1}%
\providecommand \enquote  [1]{``#1''}%
\providecommand \bibnamefont  [1]{#1}%
\providecommand \bibfnamefont [1]{#1}%
\providecommand \citenamefont [1]{#1}%
\providecommand \href@noop [0]{\@secondoftwo}%
\providecommand \href [0]{\begingroup \@sanitize@url \@href}%
\providecommand \@href[1]{\@@startlink{#1}\@@href}%
\providecommand \@@href[1]{\endgroup#1\@@endlink}%
\providecommand \@sanitize@url [0]{\catcode `\\12\catcode `\$12\catcode
  `\&12\catcode `\#12\catcode `\^12\catcode `\_12\catcode `\%12\relax}%
\providecommand \@@startlink[1]{}%
\providecommand \@@endlink[0]{}%
\providecommand \url  [0]{\begingroup\@sanitize@url \@url }%
\providecommand \@url [1]{\endgroup\@href {#1}{\urlprefix }}%
\providecommand \urlprefix  [0]{URL }%
\providecommand \Eprint [0]{\href }%
\providecommand \doibase [0]{https://doi.org/}%
\providecommand \selectlanguage [0]{\@gobble}%
\providecommand \bibinfo  [0]{\@secondoftwo}%
\providecommand \bibfield  [0]{\@secondoftwo}%
\providecommand \translation [1]{[#1]}%
\providecommand \BibitemOpen [0]{}%
\providecommand \bibitemStop [0]{}%
\providecommand \bibitemNoStop [0]{.\EOS\space}%
\providecommand \EOS [0]{\spacefactor3000\relax}%
\providecommand \BibitemShut  [1]{\csname bibitem#1\endcsname}%
\let\auto@bib@innerbib\@empty
\bibitem [{\citenamefont {Hawking}(1975)}]{Hawking74}%
  \BibitemOpen
  \bibfield  {author} {\bibinfo {author} {\bibfnamefont {S.~W.}\ \bibnamefont
  {Hawking}},\ }\bibfield  {title} {\emph {\bibinfo {title} {{Particle Creation
  by Black Holes}}},\ }\href {https://doi.org/10.1007/BF02345020} {\bibfield
  {journal} {\bibinfo  {journal} {Commun. Math. Phys.}\ }\textbf {\bibinfo
  {volume} {43}},\ \bibinfo {pages} {199} (\bibinfo {year} {1975})}\BibitemShut
  {NoStop}%
\bibitem [{\citenamefont {Bekenstein}(1973)}]{Bekenstein73}%
  \BibitemOpen
  \bibfield  {author} {\bibinfo {author} {\bibfnamefont {J.~D.}\ \bibnamefont
  {Bekenstein}},\ }\bibfield  {title} {\emph {\bibinfo {title} {Black holes and
  entropy}},\ }\href {https://doi.org/10.1103/PhysRevD.7.2333} {\bibfield
  {journal} {\bibinfo  {journal} {Phys. Rev. D}\ }\textbf {\bibinfo {volume}
  {7}},\ \bibinfo {pages} {2333} (\bibinfo {year} {1973})}\BibitemShut
  {NoStop}%
\bibitem [{\citenamefont {{Sachdev}}\ and\ \citenamefont {{Ye}}(1993)}]{SY}%
  \BibitemOpen
  \bibfield  {author} {\bibinfo {author} {\bibfnamefont {S.}~\bibnamefont
  {{Sachdev}}}\ and\ \bibinfo {author} {\bibfnamefont {J.}~\bibnamefont
  {{Ye}}},\ }\bibfield  {title} {\emph {\bibinfo {title} {{Gapless spin-fluid
  ground state in a random quantum Heisenberg magnet}}},\ }\href
  {https://doi.org/10.1103/PhysRevLett.70.3339} {\bibfield  {journal} {\bibinfo
   {journal} {Phys. Rev. Lett.}\ }\textbf {\bibinfo {volume} {70}},\ \bibinfo
  {pages} {3339} (\bibinfo {year} {1993})},\ \Eprint
  {https://arxiv.org/abs/cond-mat/9212030} {arXiv:cond-mat/9212030 [cond-mat]}
  \BibitemShut {NoStop}%
\bibitem [{\citenamefont {Kitaev}(2015)}]{kitaev_talk}%
  \BibitemOpen
  \bibfield  {author} {\bibinfo {author} {\bibfnamefont {A.}~\bibnamefont
  {Kitaev}},\ }\bibfield  {title} {\emph {\bibinfo {title} {{A simple model of
  quantum holography, talk given at KITP program: entanglement in
  Strongly-Correlated Quantum Matter}}},\ }\href@noop {} {\bibfield  {journal}
  {\bibinfo  {journal} {University of California, Santa Barbara}\ } (\bibinfo
  {year} {2015})}\BibitemShut {NoStop}%
\bibitem [{\citenamefont {Sachdev}(2015)}]{Sachdev15}%
  \BibitemOpen
  \bibfield  {author} {\bibinfo {author} {\bibfnamefont {S.}~\bibnamefont
  {Sachdev}},\ }\bibfield  {title} {\emph {\bibinfo {title}
  {{Bekenstein-Hawking Entropy and Strange Metals}}},\ }\href
  {https://doi.org/10.1103/PhysRevX.5.041025} {\bibfield  {journal} {\bibinfo
  {journal} {Phys. Rev. X}\ }\textbf {\bibinfo {volume} {5}},\ \bibinfo {pages}
  {041025} (\bibinfo {year} {2015})},\ \Eprint
  {https://arxiv.org/abs/1506.05111} {arXiv:1506.05111 [hep-th]} \BibitemShut
  {NoStop}%
\bibitem [{\citenamefont {Kitaev}\ and\ \citenamefont {Suh}(2018)}]{kitaevsuh}%
  \BibitemOpen
  \bibfield  {author} {\bibinfo {author} {\bibfnamefont {A.}~\bibnamefont
  {Kitaev}}\ and\ \bibinfo {author} {\bibfnamefont {S.~J.}\ \bibnamefont
  {Suh}},\ }\bibfield  {title} {\emph {\bibinfo {title} {{The soft mode in the
  Sachdev-Ye-Kitaev model and its gravity dual}}},\ }\href
  {https://doi.org/10.1007/JHEP05(2018)183} {\bibfield  {journal} {\bibinfo
  {journal} {Journal of High Energy Physics}\ }\textbf {\bibinfo {volume}
  {05}},\ \bibinfo {pages} {183} (\bibinfo {year} {2018})},\ \Eprint
  {https://arxiv.org/abs/1711.08467} {arXiv:1711.08467 [hep-th]} \BibitemShut
  {NoStop}%
\bibitem [{\citenamefont {Maldacena}\ and\ \citenamefont
  {Stanford}(2016)}]{Maldacena_syk}%
  \BibitemOpen
  \bibfield  {author} {\bibinfo {author} {\bibfnamefont {J.}~\bibnamefont
  {Maldacena}}\ and\ \bibinfo {author} {\bibfnamefont {D.}~\bibnamefont
  {Stanford}},\ }\bibfield  {title} {\emph {\bibinfo {title} {{Remarks on the
  Sachdev-Ye-Kitaev model}}},\ }\href
  {https://doi.org/10.1103/PhysRevD.94.106002} {\bibfield  {journal} {\bibinfo
  {journal} {Phys. Rev. D}\ }\textbf {\bibinfo {volume} {94}},\ \bibinfo
  {pages} {106002} (\bibinfo {year} {2016})},\ \Eprint
  {https://arxiv.org/abs/1604.07818} {arXiv:1604.07818 [hep-th]} \BibitemShut
  {NoStop}%
\bibitem [{\citenamefont {Maldacena}\ \emph
  {et~al.}(2016{\natexlab{a}})\citenamefont {Maldacena}, \citenamefont
  {Stanford},\ and\ \citenamefont {Yang}}]{JMDS16b}%
  \BibitemOpen
  \bibfield  {author} {\bibinfo {author} {\bibfnamefont {J.}~\bibnamefont
  {Maldacena}}, \bibinfo {author} {\bibfnamefont {D.}~\bibnamefont
  {Stanford}},\ and\ \bibinfo {author} {\bibfnamefont {Z.}~\bibnamefont
  {Yang}},\ }\bibfield  {title} {\emph {\bibinfo {title} {{Conformal symmetry
  and its breaking in two dimensional Nearly Anti-de-Sitter space}}},\ }\href
  {https://doi.org/10.1093/ptep/ptw124} {\bibfield  {journal} {\bibinfo
  {journal} {Prog. Theor. Exp. Phys.}\ }\textbf {\bibinfo {volume} {2016}},\
  \bibinfo {pages} {12C104} (\bibinfo {year} {2016}{\natexlab{a}})},\ \Eprint
  {https://arxiv.org/abs/1606.01857} {arXiv:1606.01857 [hep-th]} \BibitemShut
  {NoStop}%
\bibitem [{\citenamefont {Fu}\ \emph {et~al.}(2017)\citenamefont {Fu},
  \citenamefont {Gaiotto}, \citenamefont {Maldacena},\ and\ \citenamefont
  {Sachdev}}]{Fu16}%
  \BibitemOpen
  \bibfield  {author} {\bibinfo {author} {\bibfnamefont {W.}~\bibnamefont
  {Fu}}, \bibinfo {author} {\bibfnamefont {D.}~\bibnamefont {Gaiotto}},
  \bibinfo {author} {\bibfnamefont {J.}~\bibnamefont {Maldacena}},\ and\
  \bibinfo {author} {\bibfnamefont {S.}~\bibnamefont {Sachdev}},\ }\bibfield
  {title} {\emph {\bibinfo {title} {{Supersymmetric Sachdev-Ye-Kitaev
  models}}},\ }\href {https://doi.org/10.1103/PhysRevD.95.026009} {\bibfield
  {journal} {\bibinfo  {journal} {Phys. Rev. D}\ }\textbf {\bibinfo {volume}
  {95}},\ \bibinfo {pages} {026009} (\bibinfo {year} {2017})},\ \bibinfo {note}
  {[Addendum: Phys.Rev.D 95, 069904 (2017)]},\ \Eprint
  {https://arxiv.org/abs/1610.08917} {arXiv:1610.08917 [hep-th]} \BibitemShut
  {NoStop}%
\bibitem [{\citenamefont {Cotler}\ \emph {et~al.}(2017)\citenamefont {Cotler},
  \citenamefont {Gur-Ari}, \citenamefont {Hanada}, \citenamefont {Polchinski},
  \citenamefont {Saad}, \citenamefont {Shenker}, \citenamefont {Stanford},
  \citenamefont {Streicher},\ and\ \citenamefont {Tezuka}}]{Cotler16}%
  \BibitemOpen
  \bibfield  {author} {\bibinfo {author} {\bibfnamefont {J.~S.}\ \bibnamefont
  {Cotler}}, \bibinfo {author} {\bibfnamefont {G.}~\bibnamefont {Gur-Ari}},
  \bibinfo {author} {\bibfnamefont {M.}~\bibnamefont {Hanada}}, \bibinfo
  {author} {\bibfnamefont {J.}~\bibnamefont {Polchinski}}, \bibinfo {author}
  {\bibfnamefont {P.}~\bibnamefont {Saad}}, \bibinfo {author} {\bibfnamefont
  {S.~H.}\ \bibnamefont {Shenker}}, \bibinfo {author} {\bibfnamefont
  {D.}~\bibnamefont {Stanford}}, \bibinfo {author} {\bibfnamefont
  {A.}~\bibnamefont {Streicher}},\ and\ \bibinfo {author} {\bibfnamefont
  {M.}~\bibnamefont {Tezuka}},\ }\bibfield  {title} {\emph {\bibinfo {title}
  {{Black Holes and Random Matrices}}},\ }\href
  {https://doi.org/10.1007/JHEP05(2017)118} {\bibfield  {journal} {\bibinfo
  {journal} {JHEP}\ }\textbf {\bibinfo {volume} {05}},\ \bibinfo {pages}
  {118}},\ \bibinfo {note} {[Erratum: JHEP 09, 002 (2018)]},\ \Eprint
  {https://arxiv.org/abs/1611.04650} {arXiv:1611.04650 [hep-th]} \BibitemShut
  {NoStop}%
\bibitem [{\citenamefont {Bagrets}\ \emph {et~al.}(2017)\citenamefont
  {Bagrets}, \citenamefont {Altland},\ and\ \citenamefont
  {Kamenev}}]{Bagrets17}%
  \BibitemOpen
  \bibfield  {author} {\bibinfo {author} {\bibfnamefont {D.}~\bibnamefont
  {Bagrets}}, \bibinfo {author} {\bibfnamefont {A.}~\bibnamefont {Altland}},\
  and\ \bibinfo {author} {\bibfnamefont {A.}~\bibnamefont {Kamenev}},\
  }\bibfield  {title} {\emph {\bibinfo {title} {{Power-law out of time order
  correlation functions in the SYK model}}},\ }\href
  {https://doi.org/10.1016/j.nuclphysb.2017.06.012} {\bibfield  {journal}
  {\bibinfo  {journal} {Nucl. Phys. B}\ }\textbf {\bibinfo {volume} {921}},\
  \bibinfo {pages} {727} (\bibinfo {year} {2017})},\ \Eprint
  {https://arxiv.org/abs/1702.08902} {arXiv:1702.08902 [cond-mat.str-el]}
  \BibitemShut {NoStop}%
\bibitem [{\citenamefont {Stanford}\ and\ \citenamefont
  {Witten}(2017)}]{StanfordWitten}%
  \BibitemOpen
  \bibfield  {author} {\bibinfo {author} {\bibfnamefont {D.}~\bibnamefont
  {Stanford}}\ and\ \bibinfo {author} {\bibfnamefont {E.}~\bibnamefont
  {Witten}},\ }\bibfield  {title} {\emph {\bibinfo {title} {{Fermionic
  Localization of the Schwarzian Theory}}},\ }\href
  {https://doi.org/10.1007/JHEP10(2017)008} {\bibfield  {journal} {\bibinfo
  {journal} {Journal of High Energy Physics}\ }\textbf {\bibinfo {volume}
  {10}},\ \bibinfo {pages} {008} (\bibinfo {year} {2017})},\ \Eprint
  {https://arxiv.org/abs/1703.04612} {arXiv:1703.04612 [hep-th]} \BibitemShut
  {NoStop}%
\bibitem [{\citenamefont {Moitra}\ \emph {et~al.}(2019)\citenamefont {Moitra},
  \citenamefont {Trivedi},\ and\ \citenamefont {Vishal}}]{Moitra18}%
  \BibitemOpen
  \bibfield  {author} {\bibinfo {author} {\bibfnamefont {U.}~\bibnamefont
  {Moitra}}, \bibinfo {author} {\bibfnamefont {S.~P.}\ \bibnamefont
  {Trivedi}},\ and\ \bibinfo {author} {\bibfnamefont {V.}~\bibnamefont
  {Vishal}},\ }\bibfield  {title} {\emph {\bibinfo {title} {{Extremal and
  near-extremal black holes and near-CFT$_{1}$}}},\ }\href
  {https://doi.org/10.1007/JHEP07(2019)055} {\bibfield  {journal} {\bibinfo
  {journal} {Journal of High Energy Physics}\ }\textbf {\bibinfo {volume}
  {07}},\ \bibinfo {pages} {055} (\bibinfo {year} {2019})},\ \Eprint
  {https://arxiv.org/abs/1808.08239} {arXiv:1808.08239 [hep-th]} \BibitemShut
  {NoStop}%
\bibitem [{\citenamefont {Sachdev}(2019)}]{Sachdev19}%
  \BibitemOpen
  \bibfield  {author} {\bibinfo {author} {\bibfnamefont {S.}~\bibnamefont
  {Sachdev}},\ }\bibfield  {title} {\emph {\bibinfo {title} {{Universal low
  temperature theory of charged black holes with AdS$_2$ horizons}}},\ }\href
  {https://doi.org/10.1063/1.5092726} {\bibfield  {journal} {\bibinfo
  {journal} {J. Math. Phys.}\ }\textbf {\bibinfo {volume} {60}},\ \bibinfo
  {pages} {052303} (\bibinfo {year} {2019})},\ \Eprint
  {https://arxiv.org/abs/1902.04078} {arXiv:1902.04078 [hep-th]} \BibitemShut
  {NoStop}%
\bibitem [{\citenamefont {Iliesiu}\ and\ \citenamefont
  {Turiaci}(2021)}]{luca20}%
  \BibitemOpen
  \bibfield  {author} {\bibinfo {author} {\bibfnamefont {L.~V.}\ \bibnamefont
  {Iliesiu}}\ and\ \bibinfo {author} {\bibfnamefont {G.~J.}\ \bibnamefont
  {Turiaci}},\ }\bibfield  {title} {\emph {\bibinfo {title} {{The statistical
  mechanics of near-extremal black holes}}},\ }\href
  {https://doi.org/10.1007/JHEP05(2021)145} {\bibfield  {journal} {\bibinfo
  {journal} {Journal of High Energy Physics}\ }\textbf {\bibinfo {volume}
  {05}},\ \bibinfo {pages} {145} (\bibinfo {year} {2021})},\ \Eprint
  {https://arxiv.org/abs/2003.02860} {arXiv:2003.02860 [hep-th]} \BibitemShut
  {NoStop}%
\bibitem [{\citenamefont {Heydeman}\ \emph {et~al.}(2022)\citenamefont
  {Heydeman}, \citenamefont {Iliesiu}, \citenamefont {Turiaci},\ and\
  \citenamefont {Zhao}}]{matt22}%
  \BibitemOpen
  \bibfield  {author} {\bibinfo {author} {\bibfnamefont {M.}~\bibnamefont
  {Heydeman}}, \bibinfo {author} {\bibfnamefont {L.~V.}\ \bibnamefont
  {Iliesiu}}, \bibinfo {author} {\bibfnamefont {G.~J.}\ \bibnamefont
  {Turiaci}},\ and\ \bibinfo {author} {\bibfnamefont {W.}~\bibnamefont
  {Zhao}},\ }\bibfield  {title} {\emph {\bibinfo {title} {{The statistical
  mechanics of near-BPS black holes}}},\ }\href
  {https://doi.org/10.1088/1751-8121/ac3be9} {\bibfield  {journal} {\bibinfo
  {journal} {J. Phys. A}\ }\textbf {\bibinfo {volume} {55}},\ \bibinfo {pages}
  {014004} (\bibinfo {year} {2022})},\ \Eprint
  {https://arxiv.org/abs/2011.01953} {arXiv:2011.01953 [hep-th]} \BibitemShut
  {NoStop}%
\bibitem [{\citenamefont {{Parcollet}}\ and\ \citenamefont
  {{Georges}}(1999)}]{Parcollet1}%
  \BibitemOpen
  \bibfield  {author} {\bibinfo {author} {\bibfnamefont {O.}~\bibnamefont
  {{Parcollet}}}\ and\ \bibinfo {author} {\bibfnamefont {A.}~\bibnamefont
  {{Georges}}},\ }\bibfield  {title} {\emph {\bibinfo {title}
  {{Non-Fermi-liquid regime of a doped Mott insulator}}},\ }\href
  {https://doi.org/10.1103/PhysRevB.59.5341} {\bibfield  {journal} {\bibinfo
  {journal} {Phys. Rev. B}\ }\textbf {\bibinfo {volume} {59}},\ \bibinfo
  {pages} {5341} (\bibinfo {year} {1999})},\ \Eprint
  {https://arxiv.org/abs/cond-mat/9806119} {arXiv:cond-mat/9806119
  [cond-mat.str-el]} \BibitemShut {NoStop}%
\bibitem [{\citenamefont {{Georges}}\ \emph {et~al.}(2001)\citenamefont
  {{Georges}}, \citenamefont {{Parcollet}},\ and\ \citenamefont
  {{Sachdev}}}]{GPS2}%
  \BibitemOpen
  \bibfield  {author} {\bibinfo {author} {\bibfnamefont {A.}~\bibnamefont
  {{Georges}}}, \bibinfo {author} {\bibfnamefont {O.}~\bibnamefont
  {{Parcollet}}},\ and\ \bibinfo {author} {\bibfnamefont {S.}~\bibnamefont
  {{Sachdev}}},\ }\bibfield  {title} {\emph {\bibinfo {title} {{Quantum
  fluctuations of a nearly critical Heisenberg spin glass}}},\ }\href
  {https://doi.org/10.1103/PhysRevB.63.134406} {\bibfield  {journal} {\bibinfo
  {journal} {Phys. Rev. B}\ }\textbf {\bibinfo {volume} {63}},\ \bibinfo {eid}
  {134406} (\bibinfo {year} {2001})},\ \Eprint
  {https://arxiv.org/abs/cond-mat/0009388} {arXiv:cond-mat/0009388
  [cond-mat.str-el]} \BibitemShut {NoStop}%
\bibitem [{\citenamefont {Sachdev}(2010{\natexlab{a}})}]{SS10}%
  \BibitemOpen
  \bibfield  {author} {\bibinfo {author} {\bibfnamefont {S.}~\bibnamefont
  {Sachdev}},\ }\bibfield  {title} {\emph {\bibinfo {title} {{Holographic
  metals and the fractionalized Fermi liquid}}},\ }\href
  {https://doi.org/10.1103/PhysRevLett.105.151602} {\bibfield  {journal}
  {\bibinfo  {journal} {Phys. Rev. Lett.}\ }\textbf {\bibinfo {volume} {105}},\
  \bibinfo {pages} {151602} (\bibinfo {year} {2010}{\natexlab{a}})},\ \Eprint
  {https://arxiv.org/abs/1006.3794} {arXiv:1006.3794 [hep-th]} \BibitemShut
  {NoStop}%
\bibitem [{\citenamefont {Sachdev}(2010{\natexlab{b}})}]{Sachdev:2010uj}%
  \BibitemOpen
  \bibfield  {author} {\bibinfo {author} {\bibfnamefont {S.}~\bibnamefont
  {Sachdev}},\ }\bibfield  {title} {\emph {\bibinfo {title} {{Strange metals
  and the AdS/CFT correspondence}}},\ }\href
  {https://doi.org/10.1088/1742-5468/2010/11/P11022} {\bibfield  {journal}
  {\bibinfo  {journal} {J. Stat. Mech.}\ }\textbf {\bibinfo {volume} {1011}},\
  \bibinfo {pages} {P11022} (\bibinfo {year} {2010}{\natexlab{b}})},\ \Eprint
  {https://arxiv.org/abs/1010.0682} {arXiv:1010.0682 [cond-mat.str-el]}
  \BibitemShut {NoStop}%
\bibitem [{\citenamefont {McGreevy}(2010)}]{McGreevy10}%
  \BibitemOpen
  \bibfield  {author} {\bibinfo {author} {\bibfnamefont {J.}~\bibnamefont
  {McGreevy}},\ }\bibfield  {title} {\emph {\bibinfo {title} {{In pursuit of a
  nameless metal}}},\ }\href {https://physics.aps.org/articles/v3/83}
  {\bibfield  {journal} {\bibinfo  {journal} {Physics}\ }\textbf {\bibinfo
  {volume} {3}},\ \bibinfo {pages} {83} (\bibinfo {year} {2010})}\BibitemShut
  {NoStop}%
\bibitem [{\citenamefont {Sachdev}(1999)}]{ssbook}%
  \BibitemOpen
  \bibfield  {author} {\bibinfo {author} {\bibfnamefont {S.}~\bibnamefont
  {Sachdev}},\ }\href@noop {} {\emph {\bibinfo {title} {{Quantum Phase
  Transitions}}}},\ \bibinfo {edition} {1st}\ ed.\ (\bibinfo  {publisher}
  {Cambridge University Press},\ \bibinfo {address} {Cambridge, UK},\ \bibinfo
  {year} {1999})\BibitemShut {NoStop}%
\bibitem [{\citenamefont {{Grissonnanche}}\ \emph {et~al.}(2021)\citenamefont
  {{Grissonnanche}}, \citenamefont {{Fang}}, \citenamefont {{Legros}},
  \citenamefont {{Verret}}, \citenamefont {{Lalibert{\'e}}}, \citenamefont
  {{Collignon}}, \citenamefont {{Zhou}}, \citenamefont {{Graf}}, \citenamefont
  {{Goddard}}, \citenamefont {{Taillefer}},\ and\ \citenamefont
  {{Ramshaw}}}]{admr20}%
  \BibitemOpen
  \bibfield  {author} {\bibinfo {author} {\bibfnamefont {G.}~\bibnamefont
  {{Grissonnanche}}}, \bibinfo {author} {\bibfnamefont {Y.}~\bibnamefont
  {{Fang}}}, \bibinfo {author} {\bibfnamefont {A.}~\bibnamefont {{Legros}}},
  \bibinfo {author} {\bibfnamefont {S.}~\bibnamefont {{Verret}}}, \bibinfo
  {author} {\bibfnamefont {F.}~\bibnamefont {{Lalibert{\'e}}}}, \bibinfo
  {author} {\bibfnamefont {C.}~\bibnamefont {{Collignon}}}, \bibinfo {author}
  {\bibfnamefont {J.}~\bibnamefont {{Zhou}}}, \bibinfo {author} {\bibfnamefont
  {D.}~\bibnamefont {{Graf}}}, \bibinfo {author} {\bibfnamefont {P.~A.}\
  \bibnamefont {{Goddard}}}, \bibinfo {author} {\bibfnamefont {L.}~\bibnamefont
  {{Taillefer}}},\ and\ \bibinfo {author} {\bibfnamefont {B.~J.}\ \bibnamefont
  {{Ramshaw}}},\ }\bibfield  {title} {\emph {\bibinfo {title} {{Linear-in
  temperature resistivity from an isotropic Planckian scattering rate}}},\
  }\href {https://doi.org/10.1038/s41586-021-03697-8} {\bibfield  {journal}
  {\bibinfo  {journal} {Nature}\ }\textbf {\bibinfo {volume} {595}},\ \bibinfo
  {pages} {667} (\bibinfo {year} {2021})},\ \Eprint
  {https://arxiv.org/abs/2011.13054} {arXiv:2011.13054 [cond-mat.str-el]}
  \BibitemShut {NoStop}%
\bibitem [{\citenamefont {Hartnoll}\ and\ \citenamefont
  {Mackenzie}(2021)}]{Hartnoll21}%
  \BibitemOpen
  \bibfield  {author} {\bibinfo {author} {\bibfnamefont {S.~A.}\ \bibnamefont
  {Hartnoll}}\ and\ \bibinfo {author} {\bibfnamefont {A.~P.}\ \bibnamefont
  {Mackenzie}},\ }\bibfield  {title} {\emph {\bibinfo {title} {{Planckian
  Dissipation in Metals}}},\ }\href@noop {} {\  (\bibinfo {year} {2021})},\
  \Eprint {https://arxiv.org/abs/2107.07802} {arXiv:2107.07802
  [cond-mat.str-el]} \BibitemShut {NoStop}%
\bibitem [{\citenamefont {Gibbons}\ and\ \citenamefont
  {Hawking}(1977)}]{Gibbons_Hawking}%
  \BibitemOpen
  \bibfield  {author} {\bibinfo {author} {\bibfnamefont {G.~W.}\ \bibnamefont
  {Gibbons}}\ and\ \bibinfo {author} {\bibfnamefont {S.~W.}\ \bibnamefont
  {Hawking}},\ }\bibfield  {title} {\emph {\bibinfo {title} {Action integrals
  and partition functions in quantum gravity}},\ }\href
  {https://doi.org/10.1103/PhysRevD.15.2752} {\bibfield  {journal} {\bibinfo
  {journal} {Phys. Rev. D}\ }\textbf {\bibinfo {volume} {15}},\ \bibinfo
  {pages} {2752} (\bibinfo {year} {1977})}\BibitemShut {NoStop}%
\bibitem [{\citenamefont {'t~Hooft}(1993)}]{tHooft}%
  \BibitemOpen
  \bibfield  {author} {\bibinfo {author} {\bibfnamefont {G.}~\bibnamefont
  {'t~Hooft}},\ }\bibfield  {title} {\emph {\bibinfo {title} {{Dimensional
  reduction in quantum gravity}}},\ }\href@noop {} {\bibfield  {journal}
  {\bibinfo  {journal} {Conf. Proc. C}\ }\textbf {\bibinfo {volume} {930308}},\
  \bibinfo {pages} {284} (\bibinfo {year} {1993})},\ \Eprint
  {https://arxiv.org/abs/gr-qc/9310026} {arXiv:gr-qc/9310026} \BibitemShut
  {NoStop}%
\bibitem [{\citenamefont {Susskind}\ \emph {et~al.}(1993)\citenamefont
  {Susskind}, \citenamefont {Thorlacius},\ and\ \citenamefont
  {Uglum}}]{Susskind}%
  \BibitemOpen
  \bibfield  {author} {\bibinfo {author} {\bibfnamefont {L.}~\bibnamefont
  {Susskind}}, \bibinfo {author} {\bibfnamefont {L.}~\bibnamefont
  {Thorlacius}},\ and\ \bibinfo {author} {\bibfnamefont {J.}~\bibnamefont
  {Uglum}},\ }\bibfield  {title} {\emph {\bibinfo {title} {{The Stretched
  horizon and black hole complementarity}}},\ }\href
  {https://doi.org/10.1103/PhysRevD.48.3743} {\bibfield  {journal} {\bibinfo
  {journal} {Phys. Rev. D}\ }\textbf {\bibinfo {volume} {48}},\ \bibinfo
  {pages} {3743} (\bibinfo {year} {1993})},\ \Eprint
  {https://arxiv.org/abs/hep-th/9306069} {arXiv:hep-th/9306069} \BibitemShut
  {NoStop}%
\bibitem [{\citenamefont {Maldacena}(1998)}]{Maldacena}%
  \BibitemOpen
  \bibfield  {author} {\bibinfo {author} {\bibfnamefont {J.~M.}\ \bibnamefont
  {Maldacena}},\ }\bibfield  {title} {\emph {\bibinfo {title} {{The Large $N$
  limit of superconformal field theories and supergravity}}},\ }\href
  {https://doi.org/10.1023/A:1026654312961} {\bibfield  {journal} {\bibinfo
  {journal} {Adv. Theor. Math. Phys.}\ }\textbf {\bibinfo {volume} {2}},\
  \bibinfo {pages} {231} (\bibinfo {year} {1998})},\ \Eprint
  {https://arxiv.org/abs/hep-th/9711200} {arXiv:hep-th/9711200} \BibitemShut
  {NoStop}%
\bibitem [{\citenamefont {David}\ \emph {et~al.}(2002)\citenamefont {David},
  \citenamefont {Mandal},\ and\ \citenamefont {Wadia}}]{David02}%
  \BibitemOpen
  \bibfield  {author} {\bibinfo {author} {\bibfnamefont {J.~R.}\ \bibnamefont
  {David}}, \bibinfo {author} {\bibfnamefont {G.}~\bibnamefont {Mandal}},\ and\
  \bibinfo {author} {\bibfnamefont {S.~R.}\ \bibnamefont {Wadia}},\ }\bibfield
  {title} {\emph {\bibinfo {title} {{Microscopic formulation of black holes in
  string theory}}},\ }\href {https://doi.org/10.1016/S0370-1573(02)00271-5}
  {\bibfield  {journal} {\bibinfo  {journal} {Physics Reports}\ }\textbf
  {\bibinfo {volume} {369}},\ \bibinfo {pages} {549} (\bibinfo {year}
  {2002})},\ \Eprint {https://arxiv.org/abs/hep-th/0203048}
  {arXiv:hep-th/0203048} \BibitemShut {NoStop}%
\bibitem [{\citenamefont {Strominger}\ and\ \citenamefont
  {Vafa}(1996)}]{Strominger96}%
  \BibitemOpen
  \bibfield  {author} {\bibinfo {author} {\bibfnamefont {A.}~\bibnamefont
  {Strominger}}\ and\ \bibinfo {author} {\bibfnamefont {C.}~\bibnamefont
  {Vafa}},\ }\bibfield  {title} {\emph {\bibinfo {title} {{Microscopic origin
  of the Bekenstein-Hawking entropy}}},\ }\href
  {https://doi.org/10.1016/0370-2693(96)00345-0} {\bibfield  {journal}
  {\bibinfo  {journal} {Phys. Lett. B}\ }\textbf {\bibinfo {volume} {379}},\
  \bibinfo {pages} {99} (\bibinfo {year} {1996})},\ \Eprint
  {https://arxiv.org/abs/hep-th/9601029} {arXiv:hep-th/9601029} \BibitemShut
  {NoStop}%
\bibitem [{\citenamefont {Iliesiu}\ \emph
  {et~al.}(2022{\natexlab{a}})\citenamefont {Iliesiu}, \citenamefont {Murthy},\
  and\ \citenamefont {Turiaci}}]{luca22b}%
  \BibitemOpen
  \bibfield  {author} {\bibinfo {author} {\bibfnamefont {L.~V.}\ \bibnamefont
  {Iliesiu}}, \bibinfo {author} {\bibfnamefont {S.}~\bibnamefont {Murthy}},\
  and\ \bibinfo {author} {\bibfnamefont {G.~J.}\ \bibnamefont {Turiaci}},\
  }\bibfield  {title} {\emph {\bibinfo {title} {{Black hole microstate counting
  from the gravitational path integral}}},\ }\href@noop {} {\  (\bibinfo {year}
  {2022}{\natexlab{a}})},\ \Eprint {https://arxiv.org/abs/2209.13602}
  {arXiv:2209.13602 [hep-th]} \BibitemShut {NoStop}%
\bibitem [{\citenamefont {Vishveshwara}(1970)}]{cvc}%
  \BibitemOpen
  \bibfield  {author} {\bibinfo {author} {\bibfnamefont {C.~V.}\ \bibnamefont
  {Vishveshwara}},\ }\bibfield  {title} {\emph {\bibinfo {title} {{Scattering
  of Gravitational Radiation by a Schwarzschild Black-hole}}},\ }\href
  {https://doi.org/10.1038/227936a0} {\bibfield  {journal} {\bibinfo  {journal}
  {Nature}\ }\textbf {\bibinfo {volume} {227}},\ \bibinfo {pages} {936}
  (\bibinfo {year} {1970})}\BibitemShut {NoStop}%
\bibitem [{\citenamefont {Carullo}\ \emph {et~al.}(2021)\citenamefont
  {Carullo}, \citenamefont {Laghi}, \citenamefont {Veitch},\ and\ \citenamefont
  {Del~Pozzo}}]{BHbound}%
  \BibitemOpen
  \bibfield  {author} {\bibinfo {author} {\bibfnamefont {G.}~\bibnamefont
  {Carullo}}, \bibinfo {author} {\bibfnamefont {D.}~\bibnamefont {Laghi}},
  \bibinfo {author} {\bibfnamefont {J.}~\bibnamefont {Veitch}},\ and\ \bibinfo
  {author} {\bibfnamefont {W.}~\bibnamefont {Del~Pozzo}},\ }\bibfield  {title}
  {\emph {\bibinfo {title} {{Bekenstein-Hod Universal Bound on Information
  Emission Rate Is Obeyed by LIGO-Virgo Binary Black Hole Remnants}}},\ }\href
  {https://doi.org/10.1103/PhysRevLett.126.161102} {\bibfield  {journal}
  {\bibinfo  {journal} {Phys. Rev. Lett.}\ }\textbf {\bibinfo {volume} {126}},\
  \bibinfo {pages} {161102} (\bibinfo {year} {2021})}\BibitemShut {NoStop}%
\bibitem [{\citenamefont {Chowdhury}\ \emph {et~al.}(2022)\citenamefont
  {Chowdhury}, \citenamefont {Georges}, \citenamefont {Parcollet},\ and\
  \citenamefont {Sachdev}}]{SYKRMP}%
  \BibitemOpen
  \bibfield  {author} {\bibinfo {author} {\bibfnamefont {D.}~\bibnamefont
  {Chowdhury}}, \bibinfo {author} {\bibfnamefont {A.}~\bibnamefont {Georges}},
  \bibinfo {author} {\bibfnamefont {O.}~\bibnamefont {Parcollet}},\ and\
  \bibinfo {author} {\bibfnamefont {S.}~\bibnamefont {Sachdev}},\ }\bibfield
  {title} {\emph {\bibinfo {title} {{Sachdev-Ye-Kitaev models and beyond:
  Window into non-Fermi liquids}}},\ }\href
  {https://doi.org/10.1103/RevModPhys.94.035004} {\bibfield  {journal}
  {\bibinfo  {journal} {Rev. Mod. Phys.}\ }\textbf {\bibinfo {volume} {94}},\
  \bibinfo {pages} {035004} (\bibinfo {year} {2022})},\ \Eprint
  {https://arxiv.org/abs/2109.05037} {arXiv:2109.05037 [cond-mat.str-el]}
  \BibitemShut {NoStop}%
\bibitem [{\citenamefont {Gu}\ \emph {et~al.}(2020)\citenamefont {Gu},
  \citenamefont {Kitaev}, \citenamefont {Sachdev},\ and\ \citenamefont
  {Tarnopolsky}}]{GKST}%
  \BibitemOpen
  \bibfield  {author} {\bibinfo {author} {\bibfnamefont {Y.}~\bibnamefont
  {Gu}}, \bibinfo {author} {\bibfnamefont {A.}~\bibnamefont {Kitaev}}, \bibinfo
  {author} {\bibfnamefont {S.}~\bibnamefont {Sachdev}},\ and\ \bibinfo {author}
  {\bibfnamefont {G.}~\bibnamefont {Tarnopolsky}},\ }\bibfield  {title} {\emph
  {\bibinfo {title} {{Notes on the complex Sachdev-Ye-Kitaev model}}},\ }\href
  {https://doi.org/10.1007/JHEP02(2020)157} {\bibfield  {journal} {\bibinfo
  {journal} {Journal of High Energy Physics}\ }\textbf {\bibinfo {volume}
  {02}},\ \bibinfo {pages} {157} (\bibinfo {year} {2020})},\ \Eprint
  {https://arxiv.org/abs/1910.14099} {arXiv:1910.14099 [hep-th]} \BibitemShut
  {NoStop}%
\bibitem [{\citenamefont {Pauling}(1935)}]{Pauling}%
  \BibitemOpen
  \bibfield  {author} {\bibinfo {author} {\bibfnamefont {L.}~\bibnamefont
  {Pauling}},\ }\bibfield  {title} {\emph {\bibinfo {title} {{The Structure and
  Entropy of Ice and of Other Crystals with Some Randomness of Atomic
  Arrangement}}},\ }\href {https://doi.org/10.1021/ja01315a102} {\bibfield
  {journal} {\bibinfo  {journal} {Journal of the American Chemical Society}\
  }\textbf {\bibinfo {volume} {57}},\ \bibinfo {pages} {2680} (\bibinfo {year}
  {1935})}\BibitemShut {NoStop}%
\bibitem [{\citenamefont {Maldacena}\ \emph
  {et~al.}(2016{\natexlab{b}})\citenamefont {Maldacena}, \citenamefont
  {Shenker},\ and\ \citenamefont {Stanford}}]{Maldacena16}%
  \BibitemOpen
  \bibfield  {author} {\bibinfo {author} {\bibfnamefont {J.}~\bibnamefont
  {Maldacena}}, \bibinfo {author} {\bibfnamefont {S.~H.}\ \bibnamefont
  {Shenker}},\ and\ \bibinfo {author} {\bibfnamefont {D.}~\bibnamefont
  {Stanford}},\ }\bibfield  {title} {\emph {\bibinfo {title} {A bound on
  chaos}},\ }\href {https://doi.org/10.1007/JHEP08(2016)106} {\bibfield
  {journal} {\bibinfo  {journal} {Journal of High Energy Physics}\ }\textbf
  {\bibinfo {volume} {2016}},\ \bibinfo {pages} {106} (\bibinfo {year}
  {2016}{\natexlab{b}})}\BibitemShut {NoStop}%
\bibitem [{\citenamefont {Sen}(2012)}]{Sen12}%
  \BibitemOpen
  \bibfield  {author} {\bibinfo {author} {\bibfnamefont {A.}~\bibnamefont
  {Sen}},\ }\bibfield  {title} {\emph {\bibinfo {title} {{Logarithmic
  Corrections to $\mathcal{N}=2$ Black Hole Entropy: An Infrared Window into
  the Microstates}}},\ }\href {https://doi.org/10.1007/s10714-012-1336-5}
  {\bibfield  {journal} {\bibinfo  {journal} {Gen. Rel. Grav.}\ }\textbf
  {\bibinfo {volume} {44}},\ \bibinfo {pages} {1207} (\bibinfo {year}
  {2012})},\ \Eprint {https://arxiv.org/abs/1108.3842} {arXiv:1108.3842
  [hep-th]} \BibitemShut {NoStop}%
\bibitem [{\citenamefont {Iliesiu}\ \emph
  {et~al.}(2022{\natexlab{b}})\citenamefont {Iliesiu}, \citenamefont {Murthy},\
  and\ \citenamefont {Turiaci}}]{luca22a}%
  \BibitemOpen
  \bibfield  {author} {\bibinfo {author} {\bibfnamefont {L.~V.}\ \bibnamefont
  {Iliesiu}}, \bibinfo {author} {\bibfnamefont {S.}~\bibnamefont {Murthy}},\
  and\ \bibinfo {author} {\bibfnamefont {G.~J.}\ \bibnamefont {Turiaci}},\
  }\bibfield  {title} {\emph {\bibinfo {title} {{Revisiting the Logarithmic
  Corrections to the Black Hole Entropy}}},\ }\href@noop {} {\  (\bibinfo
  {year} {2022}{\natexlab{b}})},\ \Eprint {https://arxiv.org/abs/2209.13608}
  {arXiv:2209.13608 [hep-th]} \BibitemShut {NoStop}%
\bibitem [{\citenamefont {Preskill}\ \emph {et~al.}(1991)\citenamefont
  {Preskill}, \citenamefont {Schwarz}, \citenamefont {Shapere}, \citenamefont
  {Trivedi},\ and\ \citenamefont {Wilczek}}]{Preskill91}%
  \BibitemOpen
  \bibfield  {author} {\bibinfo {author} {\bibfnamefont {J.}~\bibnamefont
  {Preskill}}, \bibinfo {author} {\bibfnamefont {P.}~\bibnamefont {Schwarz}},
  \bibinfo {author} {\bibfnamefont {A.}~\bibnamefont {Shapere}}, \bibinfo
  {author} {\bibfnamefont {S.}~\bibnamefont {Trivedi}},\ and\ \bibinfo {author}
  {\bibfnamefont {F.}~\bibnamefont {Wilczek}},\ }\bibfield  {title} {\emph
  {\bibinfo {title} {Limitations on the statistical description of black
  holes}},\ }\href {https://doi.org/10.1142/S0217732391002773} {\bibfield
  {journal} {\bibinfo  {journal} {Modern Physics Letters A}\ }\textbf {\bibinfo
  {volume} {06}},\ \bibinfo {pages} {2353} (\bibinfo {year}
  {1991})}\BibitemShut {NoStop}%
\bibitem [{\citenamefont {Shenker}\ and\ \citenamefont
  {Stanford}(2014)}]{Shenker13}%
  \BibitemOpen
  \bibfield  {author} {\bibinfo {author} {\bibfnamefont {S.~H.}\ \bibnamefont
  {Shenker}}\ and\ \bibinfo {author} {\bibfnamefont {D.}~\bibnamefont
  {Stanford}},\ }\bibfield  {title} {\emph {\bibinfo {title} {{Black holes and
  the butterfly effect}}},\ }\href {https://doi.org/10.1007/JHEP03(2014)067}
  {\bibfield  {journal} {\bibinfo  {journal} {Journal of High Energy Physics}\
  }\textbf {\bibinfo {volume} {03}},\ \bibinfo {pages} {067} (\bibinfo {year}
  {2014})},\ \Eprint {https://arxiv.org/abs/1306.0622} {arXiv:1306.0622
  [hep-th]} \BibitemShut {NoStop}%
\bibitem [{\citenamefont {Bohigas}\ \emph {et~al.}(1984)\citenamefont
  {Bohigas}, \citenamefont {Giannoni},\ and\ \citenamefont {Schmit}}]{Bohigas}%
  \BibitemOpen
  \bibfield  {author} {\bibinfo {author} {\bibfnamefont {O.}~\bibnamefont
  {Bohigas}}, \bibinfo {author} {\bibfnamefont {M.~J.}\ \bibnamefont
  {Giannoni}},\ and\ \bibinfo {author} {\bibfnamefont {C.}~\bibnamefont
  {Schmit}},\ }\bibfield  {title} {\emph {\bibinfo {title} {Characterization of
  chaotic quantum spectra and universality of level fluctuation laws}},\ }\href
  {https://doi.org/10.1103/PhysRevLett.52.1} {\bibfield  {journal} {\bibinfo
  {journal} {Phys. Rev. Lett.}\ }\textbf {\bibinfo {volume} {52}},\ \bibinfo
  {pages} {1} (\bibinfo {year} {1984})}\BibitemShut {NoStop}%
\bibitem [{\citenamefont {Almheiri}\ \emph {et~al.}(2021)\citenamefont
  {Almheiri}, \citenamefont {Hartman}, \citenamefont {Maldacena}, \citenamefont
  {Shaghoulian},\ and\ \citenamefont {Tajdini}}]{Almheiri20}%
  \BibitemOpen
  \bibfield  {author} {\bibinfo {author} {\bibfnamefont {A.}~\bibnamefont
  {Almheiri}}, \bibinfo {author} {\bibfnamefont {T.}~\bibnamefont {Hartman}},
  \bibinfo {author} {\bibfnamefont {J.}~\bibnamefont {Maldacena}}, \bibinfo
  {author} {\bibfnamefont {E.}~\bibnamefont {Shaghoulian}},\ and\ \bibinfo
  {author} {\bibfnamefont {A.}~\bibnamefont {Tajdini}},\ }\bibfield  {title}
  {\emph {\bibinfo {title} {{The entropy of Hawking radiation}}},\ }\href
  {https://doi.org/10.1103/RevModPhys.93.035002} {\bibfield  {journal}
  {\bibinfo  {journal} {Rev. Mod. Phys.}\ }\textbf {\bibinfo {volume} {93}},\
  \bibinfo {pages} {035002} (\bibinfo {year} {2021})},\ \Eprint
  {https://arxiv.org/abs/2006.06872} {arXiv:2006.06872 [hep-th]} \BibitemShut
  {NoStop}%
\bibitem [{\citenamefont {Bousso}\ \emph {et~al.}(2022)\citenamefont {Bousso},
  \citenamefont {Dong}, \citenamefont {Engelhardt}, \citenamefont {Faulkner},
  \citenamefont {Hartman}, \citenamefont {Shenker},\ and\ \citenamefont
  {Stanford}}]{Bousso}%
  \BibitemOpen
  \bibfield  {author} {\bibinfo {author} {\bibfnamefont {R.}~\bibnamefont
  {Bousso}}, \bibinfo {author} {\bibfnamefont {X.}~\bibnamefont {Dong}},
  \bibinfo {author} {\bibfnamefont {N.}~\bibnamefont {Engelhardt}}, \bibinfo
  {author} {\bibfnamefont {T.}~\bibnamefont {Faulkner}}, \bibinfo {author}
  {\bibfnamefont {T.}~\bibnamefont {Hartman}}, \bibinfo {author} {\bibfnamefont
  {S.~H.}\ \bibnamefont {Shenker}},\ and\ \bibinfo {author} {\bibfnamefont
  {D.}~\bibnamefont {Stanford}},\ }\bibfield  {title} {\emph {\bibinfo {title}
  {{Snowmass White Paper: Quantum Aspects of Black Holes and the Emergence of
  Spacetime}}},\ }\href@noop {} {\  (\bibinfo {year} {2022})},\ \Eprint
  {https://arxiv.org/abs/2201.03096} {arXiv:2201.03096 [hep-th]} \BibitemShut
  {NoStop}%
\bibitem [{\citenamefont {Chandra}\ \emph {et~al.}(2022)\citenamefont
  {Chandra}, \citenamefont {Collier}, \citenamefont {Hartman},\ and\
  \citenamefont {Maloney}}]{Chandra}%
  \BibitemOpen
  \bibfield  {author} {\bibinfo {author} {\bibfnamefont {J.}~\bibnamefont
  {Chandra}}, \bibinfo {author} {\bibfnamefont {S.}~\bibnamefont {Collier}},
  \bibinfo {author} {\bibfnamefont {T.}~\bibnamefont {Hartman}},\ and\ \bibinfo
  {author} {\bibfnamefont {A.}~\bibnamefont {Maloney}},\ }\bibfield  {title}
  {\emph {\bibinfo {title} {{Semiclassical 3D gravity as an average of large-c
  CFTs}}},\ }\href@noop {} {\  (\bibinfo {year} {2022})},\ \Eprint
  {https://arxiv.org/abs/2203.06511} {arXiv:2203.06511 [hep-th]} \BibitemShut
  {NoStop}%
\bibitem [{\citenamefont {Guo}\ \emph {et~al.}(2020)\citenamefont {Guo},
  \citenamefont {Gu},\ and\ \citenamefont {Sachdev}}]{guo2020}%
  \BibitemOpen
  \bibfield  {author} {\bibinfo {author} {\bibfnamefont {H.}~\bibnamefont
  {Guo}}, \bibinfo {author} {\bibfnamefont {Y.}~\bibnamefont {Gu}},\ and\
  \bibinfo {author} {\bibfnamefont {S.}~\bibnamefont {Sachdev}},\ }\bibfield
  {title} {\emph {\bibinfo {title} {{Linear in temperature resistivity in the
  limit of zero temperature from the time reparameterization soft mode}}},\
  }\href {https://doi.org/10.1016/j.aop.2020.168202} {\bibfield  {journal}
  {\bibinfo  {journal} {Annals Phys.}\ }\textbf {\bibinfo {volume} {418}},\
  \bibinfo {pages} {168202} (\bibinfo {year} {2020})},\ \Eprint
  {https://arxiv.org/abs/2004.05182} {arXiv:2004.05182 [cond-mat.str-el]}
  \BibitemShut {NoStop}%
\bibitem [{\citenamefont {{Aldape}}\ \emph {et~al.}(2020)\citenamefont
  {{Aldape}}, \citenamefont {{Cookmeyer}}, \citenamefont {{Patel}},\ and\
  \citenamefont {{Altman}}}]{Altman1}%
  \BibitemOpen
  \bibfield  {author} {\bibinfo {author} {\bibfnamefont {E.~E.}\ \bibnamefont
  {{Aldape}}}, \bibinfo {author} {\bibfnamefont {T.}~\bibnamefont
  {{Cookmeyer}}}, \bibinfo {author} {\bibfnamefont {A.~A.}\ \bibnamefont
  {{Patel}}},\ and\ \bibinfo {author} {\bibfnamefont {E.}~\bibnamefont
  {{Altman}}},\ }\bibfield  {title} {\emph {\bibinfo {title} {{Solvable Theory
  of a Strange Metal at the Breakdown of a Heavy Fermi Liquid}}},\ }\href@noop
  {} {\  (\bibinfo {year} {2020})},\ \Eprint {https://arxiv.org/abs/2012.00763}
  {arXiv:2012.00763 [cond-mat.str-el]} \BibitemShut {NoStop}%
\bibitem [{\citenamefont {Esterlis}\ \emph {et~al.}(2021)\citenamefont
  {Esterlis}, \citenamefont {Guo}, \citenamefont {Patel},\ and\ \citenamefont
  {Sachdev}}]{Patel1}%
  \BibitemOpen
  \bibfield  {author} {\bibinfo {author} {\bibfnamefont {I.}~\bibnamefont
  {Esterlis}}, \bibinfo {author} {\bibfnamefont {H.}~\bibnamefont {Guo}},
  \bibinfo {author} {\bibfnamefont {A.~A.}\ \bibnamefont {Patel}},\ and\
  \bibinfo {author} {\bibfnamefont {S.}~\bibnamefont {Sachdev}},\ }\bibfield
  {title} {\emph {\bibinfo {title} {{Large $N$ theory of critical Fermi
  surfaces}}},\ }\href {https://doi.org/10.1103/PhysRevB.103.235129} {\bibfield
   {journal} {\bibinfo  {journal} {Phys. Rev. B}\ }\textbf {\bibinfo {volume}
  {103}},\ \bibinfo {pages} {235129} (\bibinfo {year} {2021})},\ \Eprint
  {https://arxiv.org/abs/2103.08615} {arXiv:2103.08615 [cond-mat.str-el]}
  \BibitemShut {NoStop}%
\bibitem [{\citenamefont {Guo}\ \emph {et~al.}(2022)\citenamefont {Guo},
  \citenamefont {Patel}, \citenamefont {Esterlis},\ and\ \citenamefont
  {Sachdev}}]{Guo2022}%
  \BibitemOpen
  \bibfield  {author} {\bibinfo {author} {\bibfnamefont {H.}~\bibnamefont
  {Guo}}, \bibinfo {author} {\bibfnamefont {A.~A.}\ \bibnamefont {Patel}},
  \bibinfo {author} {\bibfnamefont {I.}~\bibnamefont {Esterlis}},\ and\
  \bibinfo {author} {\bibfnamefont {S.}~\bibnamefont {Sachdev}},\ }\bibfield
  {title} {\emph {\bibinfo {title} {{Large-$N$ theory of critical Fermi
  surfaces. II. Conductivity}}},\ }\href
  {https://doi.org/10.1103/PhysRevB.106.115151} {\bibfield  {journal} {\bibinfo
   {journal} {Phys. Rev. B}\ }\textbf {\bibinfo {volume} {106}},\ \bibinfo
  {pages} {115151} (\bibinfo {year} {2022})},\ \Eprint
  {https://arxiv.org/abs/2207.08841} {arXiv:2207.08841 [cond-mat.str-el]}
  \BibitemShut {NoStop}%
\bibitem [{\citenamefont {Patel}\ \emph {et~al.}(2022)\citenamefont {Patel},
  \citenamefont {Guo}, \citenamefont {Esterlis},\ and\ \citenamefont
  {Sachdev}}]{Patel2}%
  \BibitemOpen
  \bibfield  {author} {\bibinfo {author} {\bibfnamefont {A.~A.}\ \bibnamefont
  {Patel}}, \bibinfo {author} {\bibfnamefont {H.}~\bibnamefont {Guo}}, \bibinfo
  {author} {\bibfnamefont {I.}~\bibnamefont {Esterlis}},\ and\ \bibinfo
  {author} {\bibfnamefont {S.}~\bibnamefont {Sachdev}},\ }\bibfield  {title}
  {\emph {\bibinfo {title} {{Universal $T$-linear resistivity in
  two-dimensional quantum-critical metals from spatially random
  interactions}}},\ }\href@noop {} {\  (\bibinfo {year} {2022})},\ \Eprint
  {https://arxiv.org/abs/2203.04990} {arXiv:2203.04990 [cond-mat.str-el]}
  \BibitemShut {NoStop}%
\end{thebibliography}%

\end{document}